\documentclass[sigconf]{acmart} 
\AtBeginDocument{%
  }

\usepackage{booktabs}
\usepackage{diagbox}
\usepackage{multirow}
\usepackage{xcolor}

\usepackage{tabularx}

\usepackage{tikz}
\usepackage{amsmath}

\usepackage{caption}
\usepackage{subcaption}

\usepackage{algorithmic}
\usepackage[linesnumbered, vlined, ruled]{algorithm2e}

\usepackage{paralist}
\setdefaultenum{1.}{a.}{i.}{A.}

\usepackage[graphicx]{realboxes}
\usepackage{float}
\usepackage{tikz}

\usepackage{colortbl}
\usepackage{tcolorbox}
\definecolor{shadecolor}{rgb}{0.92,0.92,0.92}
\newtcolorbox{mybox}{ colframe=black,colback=gray!15,boxrule=1pt,arc=2pt,left=2pt,right=2pt,top=1pt,bottom=1pt}

\usepackage{url}

\usepackage{hyperref}

\usepackage{listings}
\definecolor{codegreen}{rgb}{0,0.6,0}
\definecolor{codegray}{rgb}{0.5,0.5,0.5}
\definecolor{codepurple}{rgb}{0.58,0,0.82}
\definecolor{backcolour}{rgb}{0.95,0.95,0.95}
\definecolor{linehl}{RGB}{230,238,255}
\lstdefinestyle{mystyle}{
 commentstyle=\color{codegreen},
 keywordstyle=\color{magenta},
 numberstyle=\tiny\color{codegray},
 stringstyle=\color{codepurple},
 basicstyle=\ttfamily\footnotesize,
 breakatwhitespace=false,         
 breaklines=true,                 
 captionpos=b,                    
 keepspaces=true,
 numbers=left,  
 showspaces=false,                
 showstringspaces=false,
 showtabs=false,                  
 tabsize=1,
 xleftmargin=\parindent,
 escapeinside={(*@}{@*)},
 frame=tb,
 framerule=0.5pt,
 rulecolor=\color{black},
}
\newcommand\YAMLcolonstyle{\color{red}\mdseries}
\newcommand\YAMLkeystyle{\color{black}\bfseries}
\newcommand\YAMLvaluestyle{\color{blue}\mdseries}

\makeatletter

\newcommand\language@yaml{yaml}

\expandafter\expandafter\expandafter\lstdefinelanguage
\expandafter{\language@yaml}
{
  keywords={true,false,null,y,n},
  keywordstyle=\ttfamily\color{darkgray}\bfseries,
  basicstyle=\ttfamily\YAMLkeystyle\footnotesize,                                 
  sensitive=false,
  comment=[l]{\#},
  morecomment=[s]{/*}{*/},
  commentstyle=\color{gray}\ttfamily,
  stringstyle=\YAMLvaluestyle\ttfamily,
  moredelim=[l][\color{orange}]{\&},
  moredelim=[l][\color{magenta}]{*},
  moredelim=**[il][\YAMLcolonstyle{:}\YAMLvaluestyle]{:},   
  morestring=[b]',
  morestring=[b]",
  literate =    {---}{{\ProcessThreeDashes}}3
                {>}{{\textcolor{red}\textgreater}}1     
                {|}{{\textcolor{red}\textbar}}1 
                {\ -\ }{{\mdseries\ -\ }}3,
}

\lst@AddToHook{EveryLine}{\ifx\lst@language\language@yaml\YAMLkeystyle\fi}
\makeatother

\newcommand\ProcessThreeDashes{\llap{\color{cyan}\mdseries-{-}-}}

\usepackage{hyperref}

\usepackage{enumitem}
\usepackage{pifont}    
\definecolor{DarkGreen}{RGB}{0,215,0}
\definecolor{amethyst}{rgb}{0.6, 0.4, 0.8}
\definecolor{DarkYellow}{RGB}{230,230,0}
\definecolor{DarkRed}{RGB}{215,0,0}

\newcommand{\greencheck}{{\color{DarkGreen}\ding{52}}}   
\newcommand{\redcross}{{\color{DarkRed}\ding{56}}}       
\newcommand{\yellowexclam}{{\color{DarkYellow}\ding{115}}} 

\newcommand{\ballnumber}[1]{\tikz[baseline=(myanchor.base)] \node[circle,fill=white,draw=black,inner sep=1pt] (myanchor) {\color{black}\bfseries\footnotesize #1};}

\newif\ifcommentcond
\commentcondtrue 

\newif\ifupdatecond
\updatecondfalse 

\newcounter{wqy} 
\newcounter{fh} 
\newcounter{mb} 
\newcounter{ev} 
\newcounter{todo}

\newif\ifparasumcond
\parasumcondfalse 

\newcommand{\ignore}[1]{}

\newcommand{\system}{\textsc{Bluezz}\xspace}

\setcopyright{acmlicensed} 
\copyrightyear{2018} 
\acmYear{2018} 
\acmDOI{XXXXXXX.XXXXXXX} 
\acmConference[Conference acronym 'XX]{Make sure to enter the correct
  conference title from your rights confirmation email}{June 03--05,
  2018}{Woodstock, NY}  
\acmISBN{978-1-4503-XXXX-X/2018/06}  

\begin{document}

\title{Reactive Peripheral Modeling for Faithful Firmware Rehosting}

\author{Qinying Wang$^{*}$, Florian Hofhammer$^{*}$, Eduard Vlad$^{*}$, Jianqiang Wang$^{\dagger}$
Marcel Busch$^{*}$, \\ Shouling Ji$^{\ddagger}$ and Mathias Payer$^{*}$}
\affiliation{%
  \institution{$^{*}$École Polytechnique Fédérale de Lausanne (EPFL)\\
  $^{\dagger}$Max Planck Institute for Security and Privacy\\
  $^{\ddagger}$Zhejiang University}
  \country{\ }
}

\renewcommand{\shortauthors}{}

\begin{abstract}
Embedded BLE stacks are ubiquitous and therefore security critical.
Rehosting enables tight control and introspection during testing but
existing approaches largely fail to reach deeper protocol states and cannot 
drive the stacks beyond early-stage initialization.
This limitation is not specific to BLE, but reflects a more
fundamental limitation in existing rehosting approaches: 
their inability to faithfully model complex peripheral semantics and dependencies.
In particular, existing work typically relies on passive approximations of 
peripheral behavior, overlooking three key aspects of peripheral semantics:
(i) the interplay among interrupts, MMIO, and DMA;
(ii) implicit state transitions within peripherals;
(iii) and interactions across multiple peripherals.

To address this challenge, we propose Reactive Peripheral Modeling (RPM), 
an abstraction that models hardware peripherals as reactive and stateful systems.
RPM captures peripheral behavior through event-condition-action semantics, 
enabling faithful representation of interrupt, MMIO, DMA 
scheduling, implicit state transitions and cross-peripheral interactions.
We implement RPM in \system{} for BLE firmware rehosting and fuzzing, 
and show that such reactive modeling is necessary to reach deep protocol states, 
including connection establishment, service discovery, and post-connection processing.
We evaluate \system{} on representative BLE stacks, including NimBLE, Zephyr, 
and Nordic SoftDevice, a closed-source commercial stack.
Across 18 targets, \system{} achieves over 2.6$\times$ the basic-block 
coverage of prior state-of-the-art rehosting approaches on average.
Unlike prior approaches that remain largely confined to advertising and 
scanning logic, \system{} reliably exercises connected BLE states and uncovers 
five previously unknown vulnerabilities that manifest 
only after connection establishment.
Finally, we show that RPM generalizes beyond BLE to other embedded firmware 
across different MCUs.
\end{abstract}

\ccsdesc[500]{Do Not Use This Code~Generate the Correct Terms for Your Paper}
\keywords{Faithful rehosting, firmware fuzzing} 


\hyphenation{firm-ware re-hosting}

\maketitle


\section{Introduction} 
Bluetooth Low Energy (BLE) powers billions of devices and underpins 
security-critical applications ranging from wearables and smartphones to 
vehicles and smart home systems \cite{bluetoothtech, bluetoothmarket}.
As these devices handle sensitive data, 
vulnerabilities in embedded BLE stacks can have serious consequences. 
Prior work demonstrated how flaws in BLE stacks can allow remote 
attackers to gain code execution using crafted 
over-the-air packets, often without physical 
access \cite{garbelini2020sweyntooth, garbelini2022braktooth, wen2020firmxray, heinze2020toothpicker, wu2024sok, wu2022formal}. 

Detecting such vulnerabilities, however, remains challenging because
BLE firmware is typically closed-source, tightly coupled with specific
hardware, and difficult to control during over-the-air BLE stack testing.
Firmware rehosting is a promising approach for such testing because
it enables controlled execution, fine-grained introspection, and
scalable input generation without requiring physical devices.
However, existing rehosting approaches fail to \emph{effectively explore
the reachable attack surface} of firmware.
\textbf{First}, existing approaches
\cite{feng2020p2im, scharnowski2022fuzzware,scharnowski2023hoedur,wang2025aidfuzzer,mera2021dice,scharnowski2025gdma,lei2024friend,bley2025protocol} 
fail to drive firmware beyond early initialization
stages. In BLE, existing systems remain largely confined to advertising
and scanning logic, and rarely reach connection establishment or
post-connection processing, where much of the security-critical protocol
handling is implemented. This leaves much of the remote attack surface
unreachable.
\textbf{Second}, they expose peripheral interfaces such as 
Memory-Mapped I/O (MMIO) values and 
interrupt schedules as generic fuzzing inputs, even though these interfaces are 
not attacker-controllable in practice. 
This can drive execution into infeasible or spurious states that do not 
correspond to realistic attack scenarios.
As a result, the effective attack surface is both inaccurately
modeled and largely unreachable, leading to misleading testing
results and missed vulnerabilities.

Importantly, this limitation is not unique to BLE. 
Rather, it stems from a more fundamental challenge in peripheral modeling for 
embedded firmware analysis.
Existing emulators commonly approximate peripherals as passive MMIO 
endpoints that simply return expected values or trigger isolated events.
Real world peripherals, however, are reactive and stateful systems whose 
behavior is driven by causal event chains across MMIO accesses, DMA transfers, 
and interrupts.
When these causal semantics are missing or violated, rehosted execution
becomes semantically invalid.
Expected hardware events may never occur,
preventing firmware from reaching deep states.
Conversely, impossible events or values may be introduced, driving firmware 
into infeasible states and causing spurious crashes.
This raises a pressing question: \emph{how can we faithfully rehost
embedded firmware while preserving the reactive hardware semantics
needed to reach attacker-relevant states?}

We identify three challenges to answer this question. 
\textbf{Challenge~1:} Modeling causal hardware events.
Peripheral behavior is not determined by MMIO accesses alone, but by
the interplay among MMIO accesses, interrupts, and DMA transfers.
\textbf{Challenge~2:} Modeling implicit peripheral state transitions.
Peripherals perform implicit state transitions driven by hardware logic
rather than explicit firmware instructions.
\textbf{Challenge~3}: Modeling cross-peripheral coordination.
Peripheral behavior often spans multiple components: timers, radios,
DMA engines, and interrupt controllers must coordinate to produce the
events observed by firmware.
Together, these complex and fundamental peripheral characteristics make it 
difficult to reconstruct the causal hardware behavior required for semantically
valid execution.

\noindent\textbf{Our solution.} To address this gap, we propose \emph{Reactive
Peripheral Modeling (RPM)}, an abstraction that models hardware
peripherals as reactive systems through \emph{event-condition-action (ECA)}
semantics. 
RPM represents peripheral semantics as a set of ECA rules derived from
hardware reference manuals. 
Each rule specifies 
the triggering \textbf{event}, the hardware \textbf{conditions} under which the 
rule applies, and the resulting \textbf{actions} performed by the peripheral. 
These elements are grounded in concrete hardware artifacts, including MMIO 
registers, interrupt lines, DMA descriptors, and, when needed, internal 
peripheral states.
This design provides two key benefits. 
First, RPM captures not only MMIO interactions, but also the causal
event chains expected by the firmware, enabling faithful
rehosting and semantically valid progression into deeper states.
Second, RPM aligns fuzzing with the realistic attack surface. Since
fuzzed inputs are propagated through hardware mechanisms rather than
injected as arbitrary MMIO values or interrupts, RPM enables testing the
firmware paths that are actually reachable by attacker-controlled wireless
packets or serial inputs.

We instantiate RPM in \system{} and use embedded BLE stacks as a 
challenging case study, since BLE requires tight coordination across multiple 
peripherals and protocol stages. 
In BLE,  the attacker-controlled inputs are over-the-air radio packets and the
target states include connection establishment, service discovery, and
post-connection processing.
To make RPM practical in \system{}, we expose events, conditions, and
actions as composable primitives, allowing users to build peripheral
models by assembling reusable building blocks. 
\system{} further automates the low-level binding between RPM rules 
and hardware artifacts: it parses the official CMSIS-SVD 
(System View Description) files to recover peripheral register layouts
and access patterns, 
and hooks firmware register accesses at runtime.
When firmware writes to registers that configure or trigger peripheral
behavior, \system{} dispatches the corresponding RPM rules to update
peripheral state, schedule DMA transfers, or fire interrupts.

\noindent\textbf{Evaluation.}
We evaluate \system{} on 18 BLE firmware samples spanning three major
BLE stacks: NimBLE, Zephyr, and Nordic SoftDevice. Notably, Nordic
SoftDevice is a commercial closed-source BLE stack. Our evaluation
shows that \system{} substantially improves deep-state coverage and
vulnerability discovery over prior rehosting approaches. Specifically,
\system{} achieves over 2.6$\times$ basic-block coverage and over
4.9$\times$ BLE stack-related basic-block coverage compared to state-of-the-art (SOTA).
As far as we know, \system{} is the first rehosting-based framework to systematically exercise 
all phases of the BLE protocol stack, including scanning, advertising, 
connection, and data exchange.

More importantly, \system{} identifies five previously unknown
vulnerabilities in post-connection states, including improper input validation and double-free bugs.
Alarmingly, these vulnerabilities can be triggered
by remote attackers using malformed over-the-air packets, leading to
denial of service or unstable controller states. 
Since they reside in widely deployed BLE stack implementations, which are
reused across multiple vendors and products, they potentially impact a large
number of devices.
We have responsibly disclosed all issues, and one has already been fixed.

Finally, we evaluate the generality of RPM beyond BLE. \system{} models
nine commonly used peripherals required by BLE firmware, most of which
are not supported by existing emulators. We further apply RPM to
Grbl~\cite{grbl}, a widely used serial-protocol firmware for Computer
Numerical Control (CNC) milling. We extend \system{} to model seven
types of peripherals from a different MCU family required by the
firmware, and fuzz the firmware through its serial interface as a
practical attack interface. The results show that RPM achieves coverage
comparable to the SOTA on Grbl after about four hours of manual effort
to adapt its ECA rules. This demonstrates that RPM is extensible and can
generalize beyond BLE firmware with manageable manual effort.

In summary, our contributions are as follows:

\begin{itemize}

  \item \textbf{Novel insights}. We identify attack-surface exploration as a key limitation of
existing firmware rehosting. We propose a new modeling method, an
  event-condition-action abstraction that models peripherals as
  reactive systems and captures causal hardware event chains, implicit
  peripheral state transitions, and cross-peripheral coordination.
  
  \item \textbf{Pratical tool.} We implement RPM in \system{}, a BLE
  firmware rehosting and fuzzing framework for comprehensive BLE stack
  testing. With RPM rules and protocol-aware input models, \system{} can
  rehost and fuzz diverse BLE stacks, including closed-source stacks
  without symbols.

  \item \textbf{Extensive evaluation.} We evaluate \system{} on 18 BLE
  firmware samples across NimBLE, Zephyr, and Nordic SoftDevice. Our
  results show that \system{} outperforms prior rehosting approaches,
  reaches connection and post-connection states, and uncovers five
  previously unknown vulnerabilities.

  \item \textbf{Open source.} We will release \system{} and a BLE
  firmware benchmark suite covering major BLE stacks, roles, and
  protocol phases to support future research.

\end{itemize}

\section{Background}




\subsection{Rehosting-based Firmware Fuzzing}

Firmware running on a microcontroller unit (MCU) relies on hardware
peripherals, such as timers, radios, and UARTs, to interact with the physical
environment and communicate with external devices.
Firmware interacts with these peripherals mainly through three mechanisms:
memory-mapped I/O (MMIO), interrupts, and direct memory access (DMA).
\emph{Through MMIO}, firmware reads and writes peripheral registers to configure
hardware behavior, issue commands, and observe hardware status.
Peripheral MMIO registers can be broadly classified into three types:
control registers, which firmware writes to configure peripheral behavior
or trigger hardware tasks;
status registers, which firmware reads to observe hardware states or event
flags; and
data registers, which carry input or output data between firmware and
peripherals.
In practice, some registers may combine control and status fields, but they
still serve these basic interaction purposes.
\emph{Through interrupts}, peripherals asynchronously notify firmware when hardware
events occur, such as timer expiration, packet reception, or transfer
completion.
\emph{Through DMA}, peripherals directly read from or write to memory buffers without
requiring the CPU to explicitly move each byte.

To enable security analysis with high introspection, firmware rehosting runs 
firmware off-device in a virtual environment, and emulate the peripherals 
to advance the firmware execution.
Existing rehosting-based fuzzing approaches typically emulate hardware at 
two different abstraction layers.
\emph{At the high-level abstraction layer (HAL)} \cite{clements2020halucinator, hofhammer2024surgeon, seidel2023forming, li2021library},
the emulator redirects calls to HAL functions to custom handlers that provide equivalent peripheral
behaviors in the virtual environment.
This approach avoids modeling low-level hardware accesses, such as DMA
transfers and MMIO accesses, with manageable manual effort. 
However, it is difficult to apply to firmware without clearly identifiable HAL
functions.
\emph{At the hardware layer} \cite{feng2020p2im, mera2021dice, johnson2021jetset, scharnowski2022fuzzware, scharnowski2023hoedur,wang2025aidfuzzer, scharnowski2025gdma}, 
the emulator models low-level hardware
mechanisms such as MMIO accesses, interrupts, and DMA transfers.
Existing approaches commonly supply values for MMIO reads or DMA
buffers using fuzzed inputs, heuristics, or symbolic execution to infer
firmware-expected values. 
They also trigger interrupts randomly,
periodically, or approximately based on firmware analysis.
This layer is more scalable than HAL-level emulation and can be
effective for firmware that depends on relatively simple peripheral
behaviors. However, it becomes less effective for firmware that relies
on complex, timing-sensitive, or interdependent peripheral semantics.
In such cases, firmware-centric models may miss hardware behaviors that
are implicit in peripherals and not directly visible from firmware
execution. As a result, execution may fail to progress into deeper
states. Moreover, exposing MMIO values, DMA buffers, or interrupt
schedules as fuzzing inputs may cause the fuzzer to explore states that
are not reachable through realistic attacker-controlled interfaces.

\subsection{Realistic Attack Surface for BLE Firmware}

We consider a nearby wireless adversary that sends crafted BLE packets
over the air to a target device.
This threat model is practical and
consistent with prior BLE attacks~\cite{garbelini2020sweyntooth,
karim2023blediff,antonioli2020bias}. 
Under this threat model, the adversary controls only protocol-level 
wireless inputs, and cannot
directly control MMIO registers, DMA descriptors, interrupt schedules,
or other internal hardware events.
and post-connection data exchange.

\begin{figure}[htbp]
  \centering
  \includegraphics[width=0.75\linewidth]{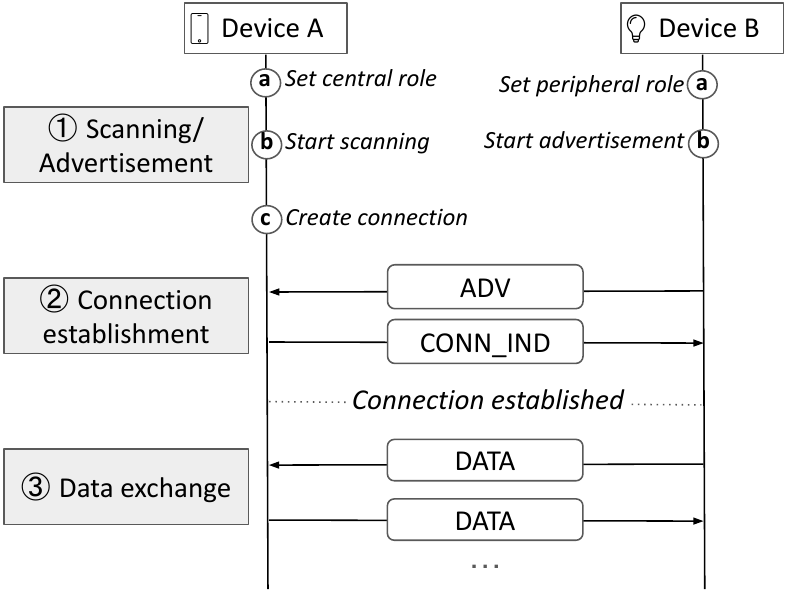} 
  \caption{Major BLE phases, including scanning/advertisement, connection 
  establishment, and data exchange.
  }
  \label{fig:usecase} 
\end{figure}

Realistic BLE firmware testing should therefore exercise firmware
through over-the-air packets. As shown in~\autoref{fig:usecase}, BLE
communication involves three main phases: \ballnumber{1}
scanning/advertising, \ballnumber{2} connection establishment, and
\ballnumber{3} data exchange. During scanning and advertising, devices
discover each other through advertisements, scan requests, and scan
responses. Connection establishment begins when a central device sends a
connection request and the peers negotiate link parameters such as
connection interval and latency.
After connection establishment, devices enter post-connection data
exchange, where higher-layer BLE logic such as service discovery and
attribute access processes attacker-controlled packets.

Thus, the realistic remote attack surface of BLE firmware is defined by
the protocol states and packet types reachable through over-the-air
communication. In particular, connection and post-connection phases are
critical because they exercise security-relevant logic beyond early
discovery. Appendix~\ref{app:blepackets} summarizes the BLE stack
layers and representative packet structures.


\section{Motivation and New Modeling Paradigm}
\label{sec:motivate}



\subsection{Why Existing Approaches Fail}
Our goal is to enable realistic attack-surface exploration while
retaining the scalability and applicability of hardware-layer
emulation, especially for firmware without clear HAL functions.
However, existing emulation approximates hardware-layer behaviors, and thus 
struggles when
firmware depends on complex, timing-sensitive, and interdependent
peripheral behavior. 

In the following, we use peripherals required by BLE firmware as
concrete examples to illustrate the behaviors that firmware
depends on and why existing hardware-layer approaches 
\cite{scharnowski2022fuzzware,scharnowski2023hoedur,feng2020p2im,scharnowski2025gdma,bley2025protocol,wang2025aidfuzzer,zhou2022your} fail to support
them faithfully. These limitations prevent rehosting systems from reaching deep
firmware states and limit realistic attack-surface exploration.

\noindent\textbf{Challenge 1. Modeling causal hardware events.}
Peripherals raise hardware events when specific conditions occur, such as data
reception, transmission completion, or timer expiration. 
In real hardware,
firmware-visible interrupts are raised only when the corresponding peripheral
event occurs and its interrupt source is enabled. The interrupt is gated by both 
the global interrupt controller, e.g., the Nested Vectored
Interrupt Controller (NVIC), and by local interrupt-enable fields in
peripheral MMIO registers.
Existing approaches often treat interrupts as generic execution-driving inputs:
once an interrupt is enabled in the NVIC, they may trigger it randomly and then supply
MMIO values on demand. 
This ignores the causal relation among peripheral events,
MMIO state, and interrupt delivery, leading to interrupts that may be
inconsistent with real hardware behavior.



Moreover, delivering an interrupt alone is insufficient to advance
firmware execution correctly. On interrupt delivery, the
interrupt service routine (ISR) expects the corresponding MMIO state
and DMA-propagated data to be consistent with that event.
For instance, in Listing~\ref{lst:radio-isr}, the RADIO interrupt line
invokes the ISR \texttt{ble\_phy\_isr()}. In Steps 1 and 2, the firmware
continues processing BLE packets only if the MMIO status registers
indicate that \texttt{EVENTS\_ADDRESS} or
\texttt{EVENTS\_DISABLED} has been raised. Moreover, the packet itself
must already have been propagated to the receive buffer via DMA.

\begin{lstlisting}[language=C,caption={Simplified Nimble BLE controller code snippet.},label={lst:radio-isr}]
/*Interrupt service routine for RADIO interrupt.*/
static void ble_phy_isr(void) 
{
  /*Read the MMIO status register for interrupt.*/
    uint32_t irq = NRF_RADIO->INTENSET;

  /*1. Check if RADIO raised EVENTS_ADDRESS */
    if ((irq & RADIO_INTENCLR_ADDRESS_Msk) &&
        NRF_RADIO->EVENTS_ADDRESS) {
    /*1.1 Process RX packets, update PHY state */
        if (ble_phy_rx_start_isr()) {
            irq &= ~RADIO_INTENCLR_DISABLED_Msk;
        }
    }
  /*2. Check if RADIO raised EVENTS_DISABLED*/
    if ((irq & RADIO_INTENCLR_DISABLED_Msk) &&
        NRF_RADIO->EVENTS_DISABLED) {
        NRF_RADIO->EVENTS_END = 0;
        NRF_RADIO->EVENTS_DISABLED = 0;
    /*2.1 check PHY state*/
        switch (g_ble_phy_data.phy_state) {
        case BLE_PHY_STATE_RX:
            if (g_ble_phy_data.phy_rx_started) {
          /*2.2 Process RX packets, update PHY state*/
                ble_phy_rx_end_isr(); 
            } 
            break;
        case BLE_PHY_STATE_TX:
          /*2.2 Process TX packets, update PHY state*/
            ble_phy_tx_end_isr();
            break;
        default:
            BLE_LL_ASSERT(0);
        }
    }
}
\end{lstlisting}

Existing rehosting approaches typically ignore the interrupt triggering 
conditions and do not model these dependencies jointly. 
Interrupt delivery, MMIO values, and DMA data are
often approximated independently, even though firmware expects them to
be mutually consistent. 
As a result, the emulator may deliver interrupts that should not occur
or fail to deliver interrupts that firmware is waiting for, wasting
fuzzing effort on infeasible execution paths. Even when an interrupt is
delivered, the handler may execute without the corresponding MMIO state
or DMA-propagated packet data. In our example, BLE firmware may read stale or
garbage data from the receive buffer, so the real attacker-controlled
packet is never processed. Consequently, the realistic attack surface
remains untested and firmware execution may fail to progress.
\begin{mybox}
  \textbf{Takeaway 1:} Trigger interrupts only when the corresponding
  peripheral event occurs and the event-specific interrupt configuration
  is enabled. Model MMIO updates, interrupt delivery, and DMA transfers
  jointly rather than independently.
\end{mybox}




\noindent\textbf{Challenge 2. Modeling implicit peripheral state transitions.}
Additionally, hardware events also have ordering dependencies. A
later event is meaningful only after earlier events have advanced the
peripheral and firmware-maintained state.

As shown in Listing~\ref{lst:radio-isr}, the firmware must first execute
the function in line 11 of Step 1.1. This handler processes
the beginning of packet reception and updates PHY state, including
whether reception has started. Only after this state transition should
the ISR be triggered again with the \texttt{EVENTS\_DISABLED}
event in Step 2, corresponding to a valid receive completion. The
firmware then calls \texttt{ble\_phy\_rx\_end\_isr()} in Step 2.2,
allowing the packet to enter BLE link-layer processing and further
advance the controller state.

Inferring these ordering and state dependencies is difficult because they
are encoded across multiple function calls, control-flow transitions, and
data-flow dependencies that jointly update the firmware-maintained
peripheral state. Existing symbolic-execution-based approaches struggle
to recover such temporal constraints. Moreover, their modeling templates
typically construct fixed value models for MMIO accesses at specific
firmware execution addresses, and therefore do not natively support
temporal constraints. As a result, they corrupt firmware-maintained
state, causing packets to be dropped, misprocessed, or never delivered to
deep BLE logic. Even worse, they may provide fresh fuzzed data whenever
the firmware reads an MMIO register or DMA buffer. This breaks packet
consistency: Steps 1 and 2 in Listing~\ref{lst:radio-isr} are expected
to process the same received packet. Consequently, the firmware no longer
observes a coherent packet reception, and the fuzzer gets stuck.

\begin{mybox}
  \textbf{Takeaway 2:} Model peripheral events as ordered state
  transitions. 
\end{mybox}


\noindent\textbf{Challenge 3. Modeling cross-peripheral coordination.}
A further challenge is that some peripheral state transitions can be
completely invisible to the firmware. For example, BLE firmware often
uses a timer to drive radio packet transmission and reception.
From the firmware code, one may
and then becomes idle. However, when the configured timer event fires, it
can directly trigger the radio to start receiving a packet without any
firmware intervention through PPI. This in turn causes a sequence of radio state
transitions and interrupts.
Such peripheral-to-peripheral coordination is supported across vendors, 
implemented in, e.g., Nordic's PPI
\cite{nordicspec}, Silicon Labs' Peripheral Reflex System
\cite{siliconspec}, and Renesas' Event Link Controller
\cite{renesasspec}.
Existing firmware-centric approaches cannot infer this hidden cross-peripheral
coordination because the key state transition does not appear as an
explicit firmware action. Consequently, they fail to trigger the IRQs
expected by the firmware and cannot drive execution into the intended BLE
logic.
\begin{mybox}
  \textbf{Takeaway 3:} Go beyond firmware-centric modeling and capture
  hardware-level cross-peripheral event propagation.
\end{mybox}

\subsection{Reactive Peripheral Modeling}
\label{sec:rpm}
Taken together, these challenges and takeaways show that modeling such
peripheral behaviors is non-trivial. Moreover, existing passive peripheral
modeling approaches are insufficient for capturing the temporal,
stateful, and cross-peripheral dependencies required by complex wireless
protocols.
This calls for a new modeling perspective, which we call
\emph{Reactive Peripheral Modeling (RPM)}. RPM treats peripherals as
stateful and reactive systems and represents their observable behaviors
through composable Event-Condition-Action (ECA) rules. We have the following
insights.

\textbf{\#1. Peripherals are stateful and reactive systems.}
A key insight behind RPM is that a peripheral should not be modeled as
a collection of independent MMIO responses or randomly/approximately
injected interrupts. Instead, it should be treated as a stateful and
reactive system whose behavior evolves as firmware issues commands to
the peripheral through control-register writes and later observes the
peripheral's responses through status registers, buffers, and interrupts.
For instance, firmware may write to a command register to start a radio
operation. The peripheral then changes its internal state, updates status
registers, fills or consumes buffers, and raises one or more interrupts
to notify firmware of progress or completion. Faithfully modeling these
behaviors enables fuzzing to exercise realistic firmware-peripheral
interactions and focus on the actual attack surface exposed by OTA
packets.


\textbf{\#2.  All observable behaviors can be captured through 
Event-Condition-Action (ECA) rules.}
We further observe that although peripheral reactions may appear complex,
their firmware-observable behaviors can be decomposed into a unified
Event-Condition-Action (ECA) abstraction. An ECA rule has the general
form: \emph{on event if condition(s) do action(s)}. In our setting, an
event captures the trigger of a peripheral reaction, a condition captures
the register-visible or implicit state that enables the reaction, and an
action captures the resulting state update or observable effect.

Events can originate from multiple sources. While firmware writes to
peripheral registers are the most direct triggers, peripheral actions
can also generate new events. For example, an action may update the
peripheral's internal state, start a follow-up task, or notify another
peripheral. These generated events may then enable subsequent ECA rules,
forming a chain of reactions.

Such composable rule chaining allows RPM to explicitly address the three
challenging behavioral patterns identified earlier.
First, it captures \emph{causal hardware events} (Challenge~1). In RPM,
an interrupt is not modeled as an independent input. Instead, it is
produced as the action of an ECA rule, whose event and condition encode
the underlying hardware cause. For example, when a packet is received
and the corresponding interrupt-enable bit is set, the rule updates
status registers, propagates data via DMA, and raises the interrupt.
This ensures that interrupts are delivered only when the corresponding
hardware event occurs and the required MMIO configurations are satisfied.
Second, it captures \emph{sequential reactions and state transitions}
(Challenge~2), where one peripheral action changes internal state or
generates a new event that enables a later reaction within the same
peripheral. This allows RPM to model delayed and multi-step peripheral
behaviors as chains of smaller reactive rules.
Third, it captures \emph{cross-peripheral dependencies} (Challenge~3),
where an action in one peripheral triggers, configures, or enables
behavior in another peripheral. This allows RPM to model interactions
among peripherals, such as timers, DMA engines, interrupt controllers,
and radio modules, without collapsing them into a monolithic model.


We further categorize and formalize the common types of events, conditions, 
and actions, as summarized in Table~\ref{tab:eca-types}.

\begin{table}[t]
  \centering
  \caption{Categorization of events, conditions, and actions in ECA modeling.}
  \label{tab:eca-types}
  \resizebox{0.9\linewidth}{!}{
  \begin{tabularx}{\columnwidth}{lX}
  \toprule
  \textbf{Component} & \textbf{Types} \\
  \midrule
  \emph{Event} & MMIO register (e.g., control, status or control-and-status registers) updates initiated by either the firmware or the peripheral. \\
  \midrule
  \multirow{2}{*}{\emph{Condition}} 
  & 1. A specific value in a peripheral register (e.g., status or control-and-status registers). \\
  & 2. An implicit peripheral state (e.g., counter values, timer expirations). \\
  \midrule
  \multirow{4}{*}{\emph{Action}} 
  & 1. Start or stop specific tasks (e.g., packet transmission/reception, timer or counter activation). \\
  & 2. Trigger an interrupt. \\
  & 3. Update specific MMIO values in peripheral registers. \\
  & 4. Modify implicit peripheral states (e.g., internal counters or states). \\
  \bottomrule
  \end{tabularx}
  }
  \end{table}

\begin{figure}[tbp]
  \centering
  \includegraphics[width=\linewidth]{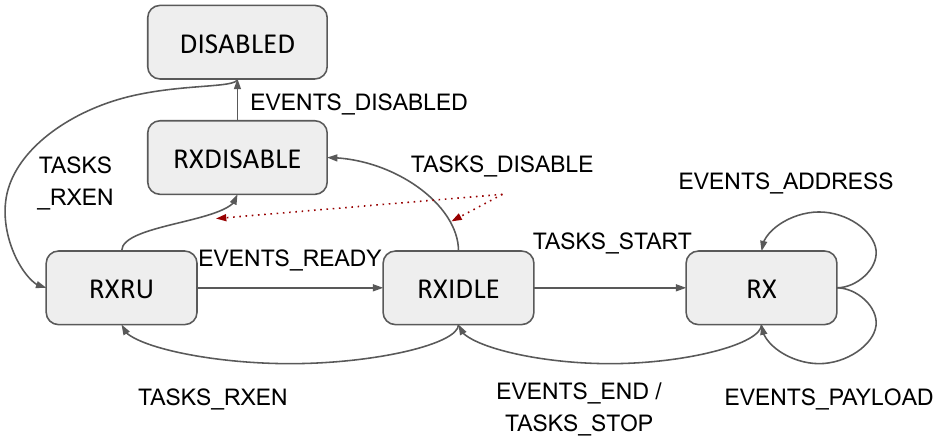} 
  \caption{The state machine of the Nordic NRF52 series RADIO peripheral. Each box 
  represents a state of the peripheral, which is reflected in the \texttt{STATE} MMIO register.
  Arrows denote state transitions triggered either by MMIO writes or by hardware events. 
  These MMIO writes target control registers typically prefixed with 
  \texttt{TASKS\_}, while hardware events are reflected in MMIO registers 
  prefixed with \texttt{EVENTS\_}. 
  The '/' symbol denotes that the state transition can be triggered by either
  a hardware event or a task command.
  }
  \label{fig:rxstatemachine} 
\end{figure}

\begin{figure*}[tbp]
  \centering
  \includegraphics[width=\linewidth]{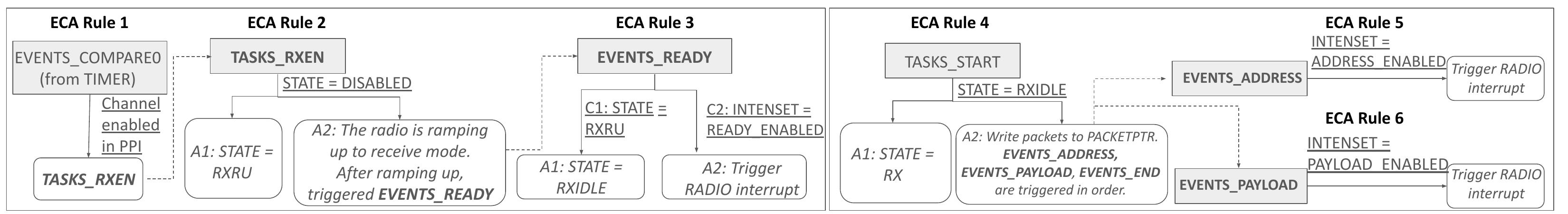} 
  \caption{Examples of the Event-Condition-Action (ECA) model for the RADIO
  peripheral. The model illustrates how the 
  peripheral transitions states based on triggers from the firmware or other 
  peripherals under conditions.
  Solid arrows denotes a trigger leading to one or multiple actions.
  A trigger may yield multiple actions when its condition holds (Rules 2 and 4), 
  or be associated with different condition–action pairs (Rule 3).
  In Rule 1, both the trigger and condition originate from other peripherals 
  (TIMER, PPI).
  In Rule 4, writing to \texttt{PACKETPTR} denotes a DMA transfer: the register 
  stores a firmware-assigned memory address, and the DMA engine writes packet 
  data to that location.
  }
  \label{fig:ecamodeling} 
\end{figure*}

\subsection{RADIO Modeling with RPM}
We use the RADIO peripheral as a concrete example to illustrate RPM.
\autoref{fig:rxstatemachine} shows the sequence of peripheral reactions
triggered when firmware configures the RADIO peripheral and issues a
command to start packet reception. \autoref{fig:ecamodeling} then shows
how RPM captures these reactions as chained ECA rules, where firmware
commands, peripheral states, and observable effects are explicitly linked.
In ECA Rule 1, a hardware event from TIMER triggers a 
RADIO task, conditioned on a status value in the PPI peripheral.
ECA rules can also be chained to represent internal peripheral progressions.
Specifically, the action of one rule may update a status register, which then 
serves as the event for a subsequent rule.
For instance, Rule 2 and Rule 3 are sequentially activated: 
in Rule 2, the RADIO transitions from \textsc{DISABLED} to \textsc{RXRU} 
when \texttt{TASKS\_RXEN} is set, 
and an action generates a hardware event \texttt{EVENTS\_READY}.
This hardware event then serves as the event in ECA Rule 3 to transition from 
\textsc{RXRU} to \textsc{RXIDLE}.
Moreover, a single rule may contain multiple actions, each encoding temporally 
ordered behavior.
For instance, in Rule 4, the packet reception process involves a series of 
steps that sequentially generate \texttt{EVENTS\_ADDRESS}, 
\texttt{EVENTS\_PAYLOAD}, and \texttt{EVENTS\_END}, reflecting the progress 
of a DMA transfer.
These hardware events, in turn, serve as ordered events for chained ECA Rules 5 and 6, 
which must be activated in sequence to faithfully reflect the 
peripheral's behaviors.

Moreover, based on these ECA rules, we further observe that events, conditions, and 
actions follow a limited set of recurring patterns, which can be instantiated 
using reusable, MMIO register–centered templates.
Together, these features demonstrate the ECA model's expressiveness in 
capturing both internal and cross-peripheral logic, including sequential 
dependencies and implicit internal state transitions.

\section{Design of \system{}}

\begin{figure*}[htbp]
  \centering
  \includegraphics[width=0.9\linewidth]{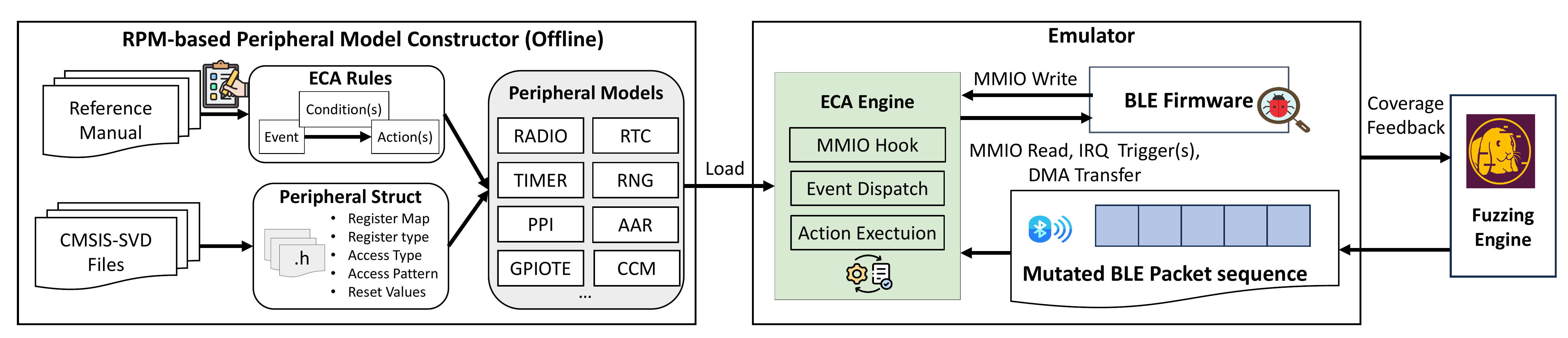} 
  \caption{ Overview of \system{}. \system{} first builds RPM-based peripheral models from hardware artifacts. The constructed models are then loaded into the emulator to facilitate firmware fuzzing with realistic attack surfaces. 
  }
  \label{fig:overview} 
\end{figure*}

\textbf{Overview.}
The previous section introduces RPM as the core abstraction for modeling
reactive peripheral behavior. In this section, we present \system{},
which instantiates RPM into a practical system for BLE firmware
rehosting and fuzzing. As shown in~\autoref{fig:overview}, \system{}
 consists of two phases: offline peripheral model construction, 
 followed by emulation and fuzzing.

The main workflow is as follows.
\system{} first builds RPM-based peripheral models from
hardware manuals and SVD files, capturing peripheral behaviors as
ECA rules. At runtime, these models are loaded into the emulator,
where firmware interactions are intercepted and translated into events.
The RPM runtime processes these events to update state, trigger
interrupts, and propagate inputs through realistic hardware semantics.
A fuzzing engine feeds inputs (e.g., BLE packets) and uses coverage
feedback to guide exploration of deeper firmware states.
\subsection{Offline Peripheral Model Construction}
\label{sec:offline}
\textbf{Inputs for peripheral model construction.}
\system{} takes two inputs, including the official reference manual and 
CMSIS-SVD files.
These artifacts provide complementary information for building peripheral models.
First, CMSIS-SVD files provide a structured and machine-readable description of
peripherals, including register layouts, address mappings, bit fields, and
access permissions.
This structured representation enables automatic parsing and serves as a
reliable foundation for constructing the skeleton of the model.
Second, the official reference manual offers detailed semantic information that
is typically absent from SVD files, such as register behaviors, side effects,
state transitions, timing constraints, and usage conditions.
These descriptions allow us to enrich the structural model with peripheral
semantics and further derive ECA rules to enable reactive peripheral modeling.
In addition, these artifacts are generally publicly available because they are
released by the hardware vendor for downstream IoT vendors and developers.
Downstream vendors rely on them to configure peripherals, port device drivers,
and develop their own firmware and applications.

\noindent\textbf{SVD-derived peripheral skeleton construction.}
\system{} first constructs peripheral skeletons from the CMSIS-SVD file. This
step is fully automatic because the SVD file provides a machine-readable
description of the MCU's peripheral layout.
For each peripheral instance, \system{} extracts its base address, register map,
register offsets relative to the peripheral base, access permissions, access patterns, and reset
values. Conceptually, this process resembles generating the macros, structures,
and register definitions used by vendor-provided header files. In \system{},
however, the register map is materialized as an emulator-side data structure,
whose fields are initialized according to the reset values specified in the SVD
file.
During emulation, firmware--peripheral interactions are represented as reads and
writes to this data structure. \system{} installs hooks for every generated
register access. On a read, the hook returns the current value of the
corresponding register to the firmware. On a write, \system{} updates the
register state by default according to the register's access type and observed
access pattern.

%

\textbf{ECA rule generation.}
This step takes the peripheral reference manual as input and enriches the
SVD-derived skeleton with ECA rules, thereby completing the peripheral model.
As summarized in Table~\ref{tab:eca-types}, an event typically corresponds to a
firmware operation on an MMIO register, such as writing a control register; a
condition is encoded as a predicate over register values or runtime state; and
an action is implemented using a fixed set of reusable execution primitives.
Thus, ECA rule generation amounts to enumerating, for each MMIO write hook, the
possible triggering events, their applicable constraints, and the side effects
that should be performed when the rule fires.
%

Although this step requires manual inspection of the reference manual, the
effort is manageable for two reasons. First, \system{} provides a small set of
reusable action primitives that cover common peripheral behaviors, such as
updating registers, starting or canceling timers, transferring data, propagating
hardware events, and injecting interrupts. As a result, analysts only need to
instantiate these primitives with peripheral-specific parameters, rather than
implementing new handlers from scratch. More details about the ECA action primitives and their semantics can be found in
\autoref{app:ecatemplates}.

Second, we propose a register-centric and side-effect-guided rule extraction
methodology. Specifically, for each peripheral, we first identify its control
registers and collect all manual descriptions related to these registers. We
then derive ECA rules by analyzing which values can be written to the register
and what behavior the peripheral performs under each value. After modeling the
direct behavior of a control register, we further inspect its side effects. If a
behavior updates another register, triggers a hardware event, starts a timer, or
injects an interrupt, we recursively construct the corresponding ECA rules for
the affected register or runtime signal. This process continues until all
reachable registers and side effects have been covered.
For example, an RNG peripheral may expose two control registers,
\texttt{TASKS\_START} and \texttt{TASKS\_STOP}. Writing \texttt{1} to these
registers starts and stops random-number generation, respectively. \system{}
models the start operation using a timer-based action primitive: after a
specified delay, the primitive writes a generated value to the peripheral's data
register and raises the corresponding event. The primitive is therefore
parameterized by the delay, destination register, and write size. Since the data
register update itself causes additional observable behavior, we further inspect
the data register and event register descriptions, and add rules that propagate
the generated hardware event. In this way, the extraction process follows the
side effects induced by each modeled behavior until the peripheral's reactive
semantics are fully captured.

\noindent\textbf{Binding ECA rules into executable handlers.}
After the semantic rules and structural layouts are prepared, \system{}
binds them together by resolving the symbolic register references appearing in
each ECA rule to concrete fields in the generated \texttt{Peripheral}
structure. This step is fully automated once the rules are specified in the
structured ECA form, as all register names can be resolved against the
SVD-derived layout and all actions correspond to predefined execution
primitives.
Concretely, events are attached to the corresponding MMIO locations or internal
signals that can trigger them; conditions are compiled into predicates over
register fields and runtime state; and actions are mapped to built-in handlers
that update registers, start timers, perform DMA-like memory transfers, route
hardware events across peripherals, or inject interrupts. Through this process,
each declarative ECA rule is translated into an executable event-driven handler.
At runtime, these handlers are indexed by the accessed register or triggering
signal. When firmware performs an MMIO access, the emulator resolves the target
peripheral instance, dispatches the associated handlers, evaluates their
conditions, and applies the resulting side effects with the correct state and
timing context. This design effectively reduces peripheral emulation to a
table-driven dispatch over register-indexed handlers, enabling efficient and
reactive execution.

\subsection{Runtime Emulation and Fuzzing}
\label{sec:runtime}
\textbf{ECA engine runtime.}
After the peripheral models are constructed offline, \system{} uses them to
rehost and fuzz BLE firmware in an emulator. At runtime, the emulator executes
the firmware binary and intercepts all MMIO accesses to modeled peripherals.
For each access, \system{} locates the corresponding peripheral instance using
the SVD-derived layout, updates the register state according to the default
access semantics, and dispatches the bound ECA handlers associated with the
accessed register or internal signal. The handlers then evaluate their
conditions and perform the corresponding actions, such as updating registers,
scheduling timers, propagating hardware events, or injecting interrupts.
This runtime design allows peripheral behavior to be modeled reactively. Instead
of emulating an entire peripheral as a monolithic device model, \system{}
executes only the rules triggered by firmware-visible events. This makes the
emulation lightweight while still preserving the causal relationship between
firmware operations and peripheral responses. In particular, delayed behaviors
are realized through emulator-managed timers, and chained behaviors are handled
by allowing actions to trigger internal events that may activate additional ECA
rules.

\noindent\textbf{Fuzzing interface and fuzzing loop.}
During fuzzing, \system{} drives the rehosted firmware with mutated BLE packets.
Because firmware--peripheral interactions are mediated by the generated
peripheral model, the values read by the firmware are produced by prior
firmware-issued commands and the resulting ECA-rule executions. This allows
\system{} to focus fuzzing on realistic external attack surfaces rather than
arbitrarily perturbing internal register state.
For BLE firmware, the primary fuzzing interface is the packet-reception path of
the radio peripheral. \system{} exposes fuzzed inputs through reusable action
primitives, such as a primitive that fetches mutated packet bytes and writes
them into the peripheral receive buffer or data register. These primitives are
invoked by ECA rules that model packet-reception behavior, allowing fuzzed BLE
packets to enter the firmware through the same peripheral-facing interface used
during real-world execution.
As the firmware processes each input, \system{} collects code coverage and crash
signals, and uses this feedback to guide subsequent mutations. In this way,
\system{} supports faithful firmware execution while focusing testing on
realistic firmware attack surfaces.




\section{Implementation}
We implement \system{} on top of Fuzzware, which itself builds on the Unicorn
engine. The ECA engine is integrated into Fuzzware's harness, allowing \system{}
to reuse Fuzzware's interrupt and Nested Vectored Interrupt Controller (NVIC)
subsystems for faithful firmware execution.
\system{} targets the Nordic nRF52840 board. For this platform, \system{}
models nine commonly used peripherals that are not fully supported by existing
firmware emulators. The implementation consists of both C and Python code. The
Python components mainly automate SVD-derived peripheral skeleton construction,
while the C components implement the runtime ECA engine, reusable action
primitives, and rule binding logic. The above core framework contains over 11K lines of code. In addition, \system{} generates over 10K lines of peripheral model code for the nRF52840 platform, among which roughly 2K lines are manually written ECA rules. 
This code-size breakdown shows that most peripheral model structure can be generated automatically from SVD files, while manual effort is primarily limited to encoding missing behavioral semantics from the reference manual.
 



\section{Evaluation}
\label{sec:evaluation}

In this section, we comprehensively evaluate \system{} and demonstrate the
effectiveness of RPM in improving the rehosting and fuzzing of embedded
BLE firmware. In particular, we aim to answer the following research
questions.

\textbf{RQ1:} How does \system{} compare with existing approaches in
fuzzing capability for vendor embedded BLE stacks?

\textbf{RQ2:} Can \system{} systematically drive embedded BLE stacks into
deeper protocol phases and code paths compared to existing fuzzers?

\textbf{RQ3:} Can \system{} discover BLE stack-related vulnerabilities
that are missed by existing approaches?

\textbf{RQ4:} Can \system{} be extended beyond BLE firmware to support
other protocol-driven embedded firmware?




\subsection{Experiment Setup.}

\textbf{Collected samples.}
We evaluate \system{} on 18 firmware samples built on three representative
real-world BLE stacks, as summarized in \autoref{tab:collected-samples}.
Specifically, we
include NimBLE, Zephyr, and Nordic SoftDevice. These
stacks are representative of both open-source and commercial BLE deployments:
NimBLE is used in Apache Mynewt and Espressif platforms, Zephyr is a widely
adopted RTOS stack for embedded devices, and SoftDevice powers many commercial
Nordic BLE SoCs used in wearables and sensors. Notably, SoftDevice is
closed-source and obfuscated, making it an important target for evaluating
\system{} beyond source-available stacks.

To test the BLE stack comprehensively, we distill five representative use cases 
from the BLE core 
specification~\cite{woolley2019bluetooth} (Table~\ref{tab:usecases} in 
Appendix~\ref{app:usecase}) and implement corresponding drivers on target BLE stacks. 
We finally have 18 firmware samples, with six samples per stack.
They collectively cover the
full range of BLE roles, including scanner, advertiser, central, peripheral,
GATT client, and GATT server, as well as major protocol functionalities such
as connection establishment, parameter negotiation, service discovery, and
L2CAP communication. For each sample, we additionally construct a
role-aligned input model that captures the expected packet formats and
interaction patterns. 
To construct these models, we collect BLE interaction traces using an 
Ubertooth One from Great Scott Gadgets, parse the captured packets with Scapy, 
and derive input models from the parsed packet sequences.
Together, these samples provide a realistic and diverse
benchmark for evaluating \system{} across BLE stacks.

\begin{table}[t]
  \caption{Real-world BLE stacks in our collected samples.}
  \label{tab:collected-samples}
  \centering
  \resizebox{\linewidth}{!}{%
  \begin{tabular}{@{}lllll@{}}
    \toprule
    Name & Version & Description & GitHub Stars & Samples \\ \midrule
    NimBLE & v1.5.0 & Open-source BLE stack & 866 & 6 \\
    Zephyr & v4.2.1 & Open-source BLE stack & 15.1k & 6 \\
    Nordic SoftDevice & v2.9.1 & Commercial BLE stack & 1.3k & 6 \\ \bottomrule
  \end{tabular}%
  }
\end{table}

\begin{table*}[htbp]
  \caption{Comparison with BLE stack and rehosting-based 
  fuzzers. 
  \greencheck{}: full support. 
  \yellowexclam{}: partial support. 
  \redcross{}: no support.}
  \label{tab:fuzzingcapbility}
  \centering
  \resizebox{0.8\textwidth}{!}{%
  \begin{tabular}{@{}lccccccc@{}}
  \toprule
  Fuzzers & \begin{tabular}[c]{@{}l@{}}Closed-source\\ target support\end{tabular} &
            \begin{tabular}[c]{@{}l@{}}Hardware\\ independent \end{tabular} &
            \begin{tabular}[c]{@{}l@{}}Universal fuzzing\\ harness \end{tabular} &
            \begin{tabular}[c]{@{}l@{}}Fully featured runtime\\ (MMIO/DMA/IRQ)\end{tabular} &
            \begin{tabular}[c]{@{}l@{}}Pratical attack\\ surface \end{tabular} &
            \begin{tabular}[c]{@{}l@{}}Coverage\\ guidance \end{tabular} &
            \begin{tabular}[c]{@{}l@{}}Bug\\ introspection \end{tabular} \\ \midrule
  BrakTooth\cite{garbelini2022braktooth}   &  \greencheck   & \redcross   & \greencheck & \greencheck   & \greencheck   & \redcross    & \redcross   \\
  BLEDiff\cite{karim2023blediff}           &  \greencheck   & \redcross   & \greencheck & \greencheck   & \greencheck   & \redcross    & \redcross   \\
  Frankenstein\cite{ruge2020frankenstein}  &  \yellowexclam & \redcross   & \redcross   & \redcross     & \greencheck   & \greencheck  & \greencheck \\
  BluEMan\cite{kao2025blueman}             &  \redcross     & \greencheck & \redcross   & \yellowexclam & \greencheck   & \greencheck  & \greencheck \\
  Fuzzware\cite{scharnowski2022fuzzware}   &  \greencheck   & \greencheck & \greencheck & \yellowexclam & \redcross     & \greencheck  & \greencheck \\
  Hoedur\cite{scharnowski2023hoedur}       &  \greencheck   & \greencheck & \greencheck & \yellowexclam & \redcross     & \greencheck  & \greencheck \\
  PEMU\cite{bley2025protocol}              &  \greencheck   & \greencheck & \greencheck & \yellowexclam & \redcross     & \greencheck  & \greencheck \\
  SEmu\cite{zhou2022your}                  &  \greencheck   & \greencheck & \greencheck & \yellowexclam & \yellowexclam & \greencheck  & \greencheck \\
  Bluezz                                   &  \greencheck   & \greencheck & \greencheck & \greencheck   & \greencheck   & \greencheck  & \greencheck \\ \bottomrule
  \end{tabular}%
  }
  \end{table*}
  
\textbf{Experiment settings.}
We perform our evaluation on a 16 core 
Intel Xeon Gold 5218 CPU @ 2.20GHz with a 64 GB RAM server running a Ubuntu 
22.04 LTS OS.
For each target, we gave each fuzzer one physical CPU core and performed the 
fuzzing ten times according to the fuzzing evaluation 
guidelines \cite{klees2018evaluating, schloegel2024sok}.


\subsection{Fuzzing Capability Comparison (RQ1)}
To characterize the trade-off between hardware-layer rehosting and faithful
peripheral modeling, we collect and analyze recent state-of-the-art studies on
BLE stack fuzzing and rehosting-based firmware fuzzing.
We compare these fuzzing tools along two dimensions: \emph{target scalability} and
\emph{practical fuzzing capability}. Target scalability captures whether a tool
can be applied to vendor BLE firmware, especially closed-source targets,
without depending on device-specific manual effort. We consider three
requirements: (\romannumeral1) \emph{closed-source target support}, meaning the
tool can fuzz proprietary BLE stacks; (\romannumeral2) \emph{hardware
independence}, meaning the tool does not rely on real hardware, memory dumps,
or externally captured device state during testing; and (\romannumeral3)
\emph{universal fuzzing harness}, meaning fuzzing new stacks does not require
manually defining stack-specific fuzz entry points or custom packet injection
logic.
Practical fuzzing capability captures whether a tool can exercise BLE firmware
under realistic attack conditions while still supporting effective analysis. We
consider four requirements: (\romannumeral1) \emph{fully featured runtime},
meaning the emulated environment provides the peripherals and interactions
needed by the BLE stack; (\romannumeral2) \emph{practical attack vector},
meaning inputs are injected as OTA BLE packets rather than emulator-specific
bytes; (\romannumeral3) \emph{coverage guidance}, meaning the fuzzer can use
runtime feedback to steer exploration; and (\romannumeral4) \emph{bug
introspection}, meaning the system supports crash triage and collects execution
context useful for root-cause analysis and vulnerability classification.

As shown in \autoref{tab:fuzzingcapbility}, we compare \system{} against eight
representative BLE firmware fuzzers and general-purpose rehosting fuzzers.
Existing BLE-specific fuzzers based on rehosting or simulation
~\cite{ruge2020frankenstein,kao2025blueman} still have limited support for
closed-source targets, stack-agnostic harness construction, and fully featured
runtime environments. General rehosting-based
fuzzers~\cite{scharnowski2022fuzzware,scharnowski2023hoedur,bley2025protocol}
typically treat injected data as a mixture of emulation control and fuzzing
input, which weakens both execution fidelity and the realism of the attack
vector. SEMu~\cite{zhou2022your} emulates peripherals from specifications, but
its current models do not capture temporal sequencing or cross-peripheral
dependencies, especially for radio-related behavior. As a result, these
approaches do not simultaneously provide the capabilities needed for
comprehensive, hardware-independent, OTA-oriented fuzzing of vendor BLE
firmware. In contrast, \system{} is designed to satisfy all of these
requirements.

Notably, \system{} trades breadth for semantic fidelity. Compared with
black-box rehosting fuzzers such as Fuzzware, \system{} requires additional
effort to model MCU-specific peripheral behaviors and protocol semantics.
However, this modeling enables more realistic attack-surface exploration:
\system{} drives firmware through over-the-air BLE inputs and preserves the
reactive interactions among peripherals. As a result, it can exercise
peripheral-intensive firmware and reach protocol states that are difficult to
trigger with generic rehosting approaches.



\subsection{Fuzzing Coverage Analysis (RQ2)}
\begin{figure*}[tbp]
  \centering
  \includegraphics[width=\linewidth]{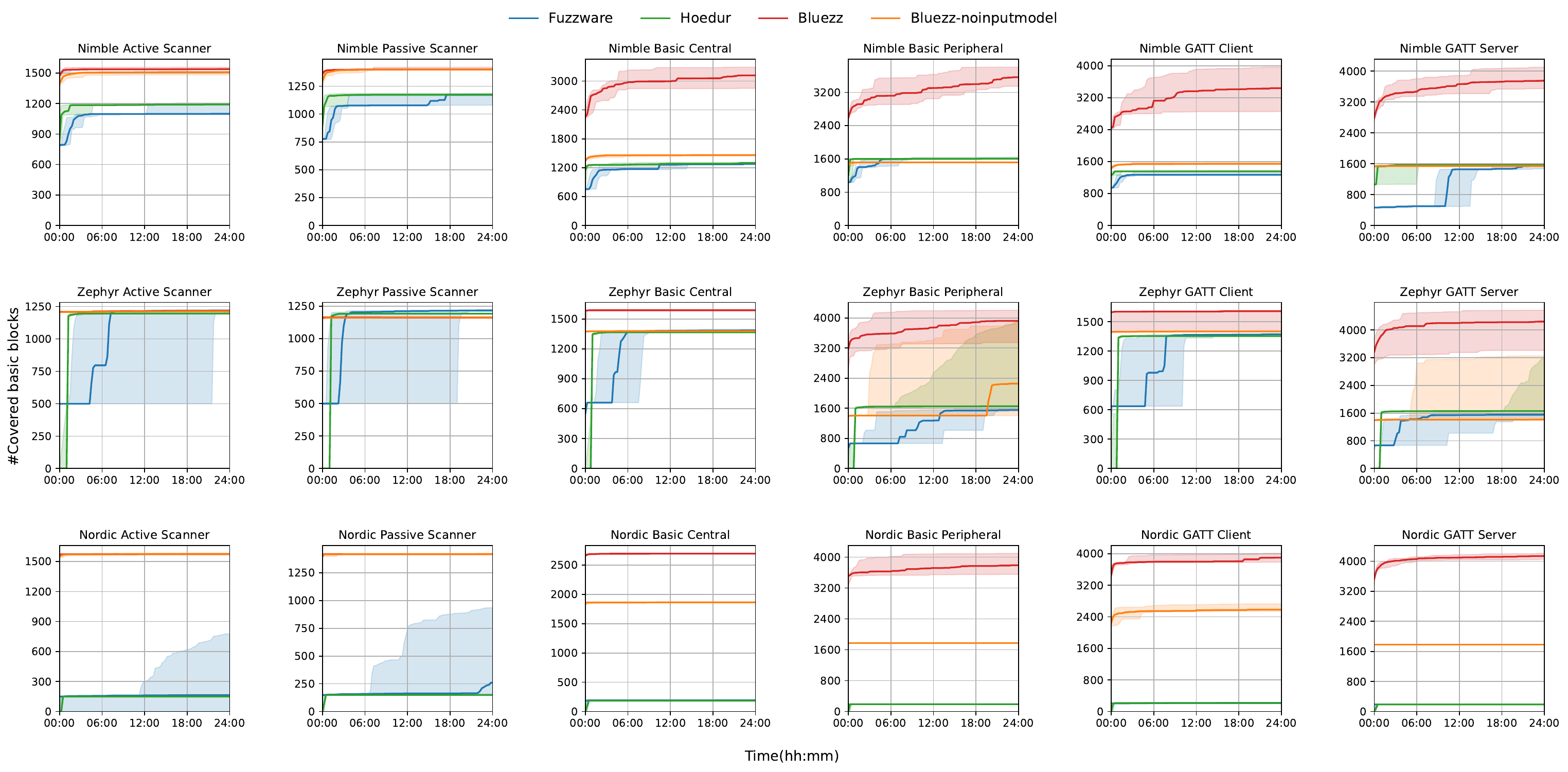} 
  \caption{ 
    Unique stack-related basic block coverage achieved by \system{}, Fuzzware, 
    and Hoedur for 18 BLE samples over 24-hour runs. Solid lines show the median; shaded areas show the minimum and maximum across 10 runs.  
  }
  \label{fig:stackcov} 
\end{figure*}

To evaluate the benefit of RPM modeling, we conduct extensive evaluation
on 18 BLE firmware samples, including basic block coverage and crash analysis.
Since existing BLE fuzzing tools either cannot target closed-source BLE stacks,
have limited scalability to new targets, or do not support coverage
calculation, we quantitatively compare \system{} with two state-of-the-art
rehosting-based fuzzers, Fuzzware~\cite{scharnowski2022fuzzware} and
Hoedur~\cite{scharnowski2023hoedur}.
Thanks to the faithful modeling, \system{} naturally supports structured BLE packets as fuzzing seeds.
We additionally evaluate \system{} without input models as an internal
baseline.
We use BLE-stack-related basic-block coverage as our primary coverage metric
because it directly measures the attack-surface exploration capability.
Total firmware coverage can therefore be misleading as Fuzzware and Hoedur may 
introduce spurious coverage that is unlikely to correspond to
realistic BLE execution paths triggered by over-the-air inputs.
To compute BLE-stack-related coverage, we identify stack code differently
depending on target availability. 
For NimBLE and Zephyr, where source code is available, we extract symbols and 
basic-block boundaries and identify BLE-stack blocks from the corresponding
host and controller modules.
In contrast, Nordic’s SoftDevice is a closed-source binary blob with obfuscated 
symbols, preventing function attribution. We therefore conservatively treat 
the entire blob as BLE-stack code and count any executed block as stack coverage.
While this may slightly overestimate BLE-specific coverage, it ensures a fair 
and consistent comparison across platforms.

\begin{figure*}[t!]
	\centering
    \begin{subfigure}{0.24\textwidth}
        \centering
        \includegraphics[width=0.8\textwidth]{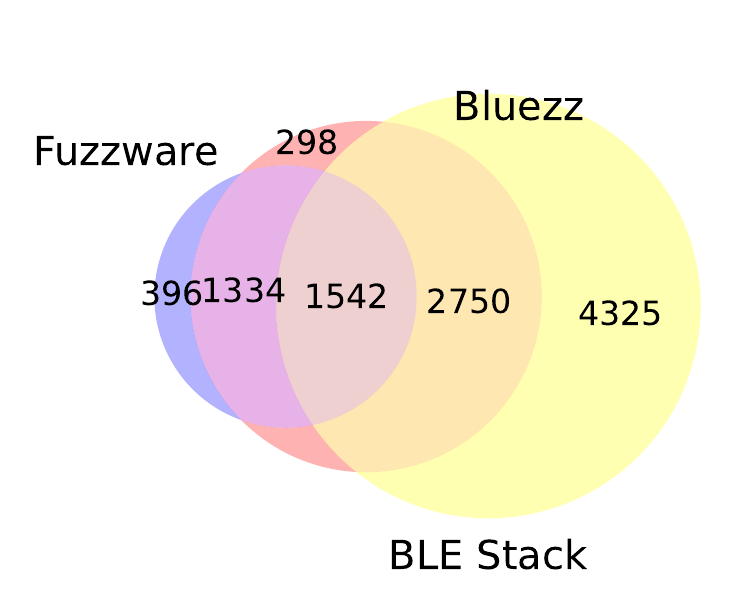}
        \caption{Nimble GATT Server}
    \end{subfigure}%
    \begin{subfigure}{0.24\textwidth}
        \centering
        \includegraphics[width=0.8\textwidth]{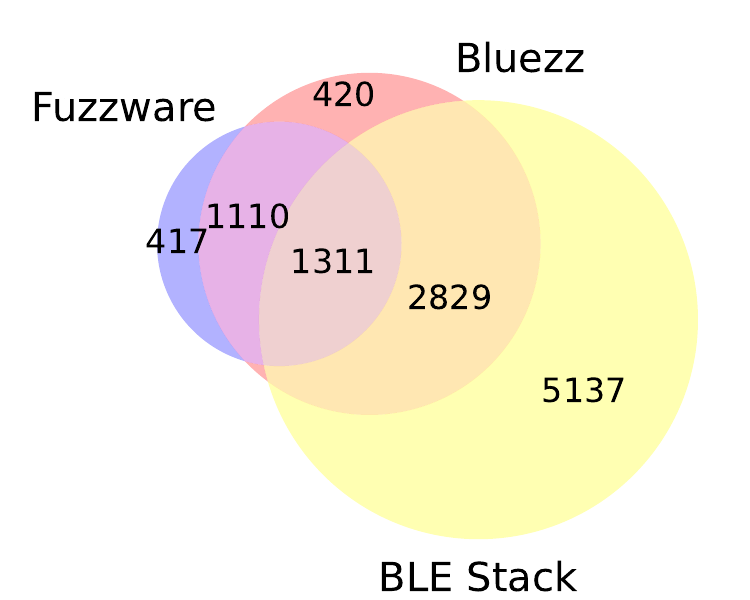}
        \caption{Nimble GATT Client}
    \end{subfigure}%
    \begin{subfigure}{0.24\textwidth}
        \centering
        \includegraphics[width=0.8\textwidth]{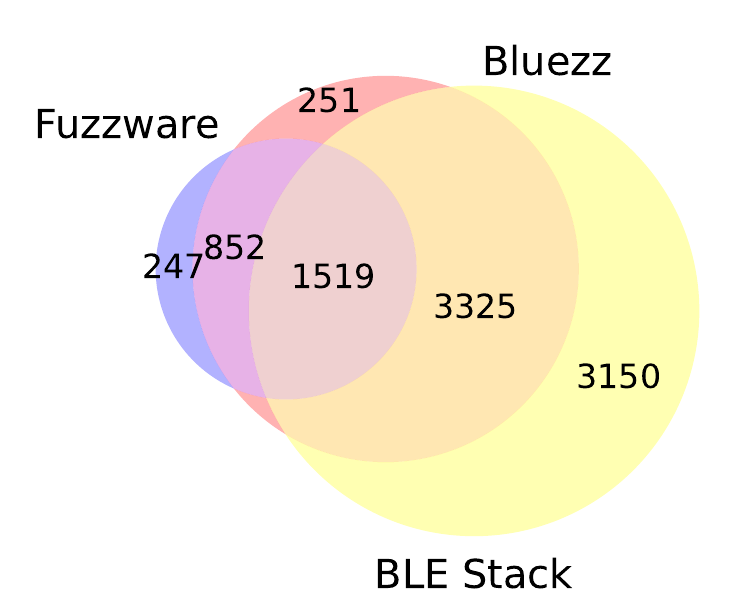}
        \caption{ Zephyr GATT Server}
    \end{subfigure}%
    \begin{subfigure}{0.24\textwidth}
        \centering
        \includegraphics[width=0.8\textwidth]{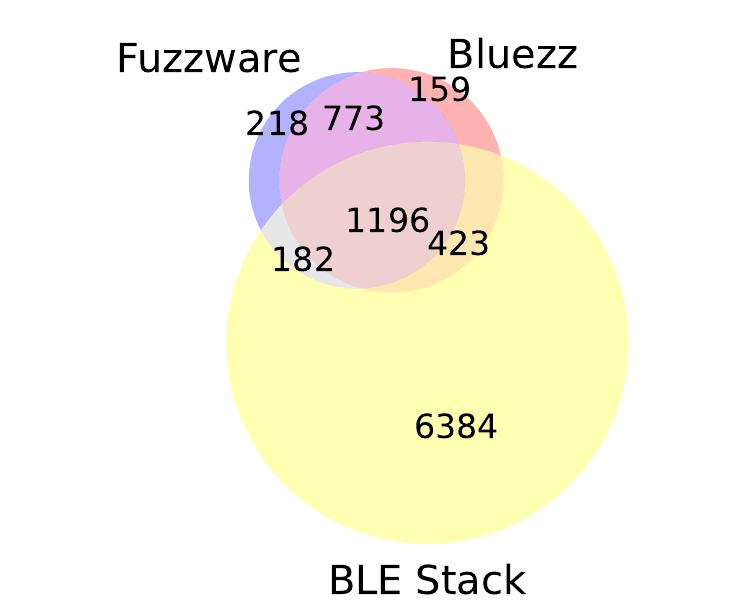}
        \caption{Zephyr GATT Client}
    \end{subfigure}
    \\
    \begin{subfigure}{0.24\textwidth}
        \centering
        \includegraphics[width=0.8\textwidth]{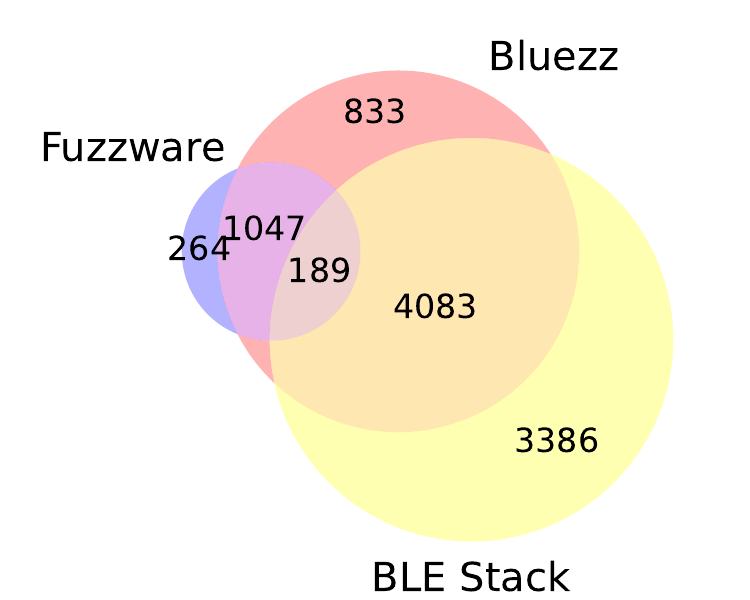}
        \caption{Nordic Basic Peripheral}
    \end{subfigure}%
    \begin{subfigure}{0.24\textwidth}
        \centering
        \includegraphics[width=0.8\textwidth]{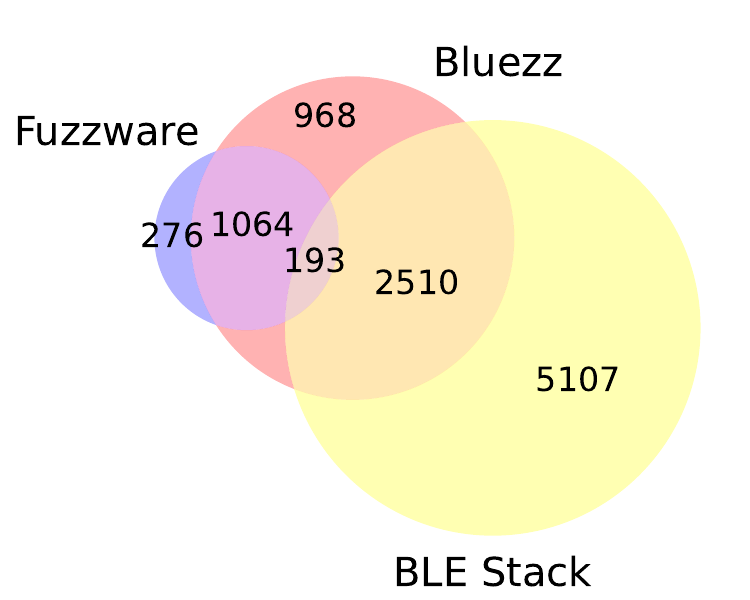}
        \caption{Nordic Basic Central}
    \end{subfigure}%
    \begin{subfigure}{0.24\textwidth}
        \centering
        \includegraphics[width=0.8\textwidth]{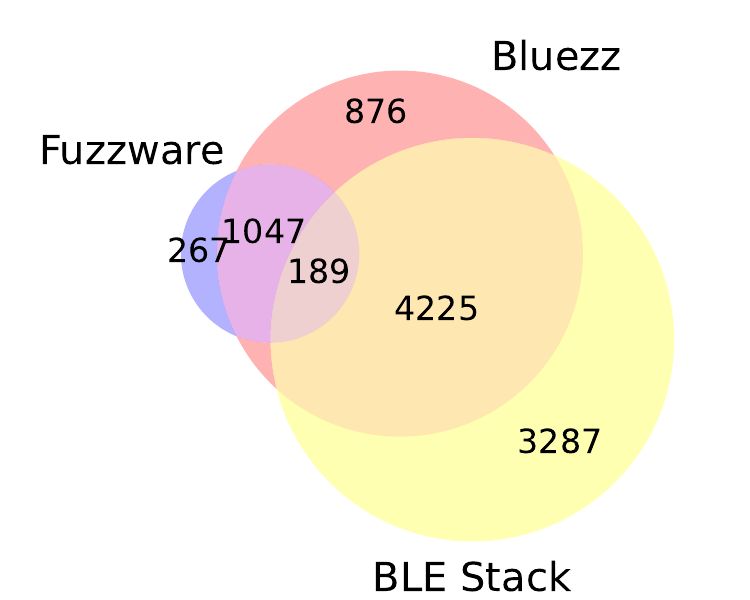}
        \caption{Nordic GATT Server}
    \end{subfigure}%
    \begin{subfigure}{0.24\textwidth}
        \centering
        \includegraphics[width=0.8\textwidth]{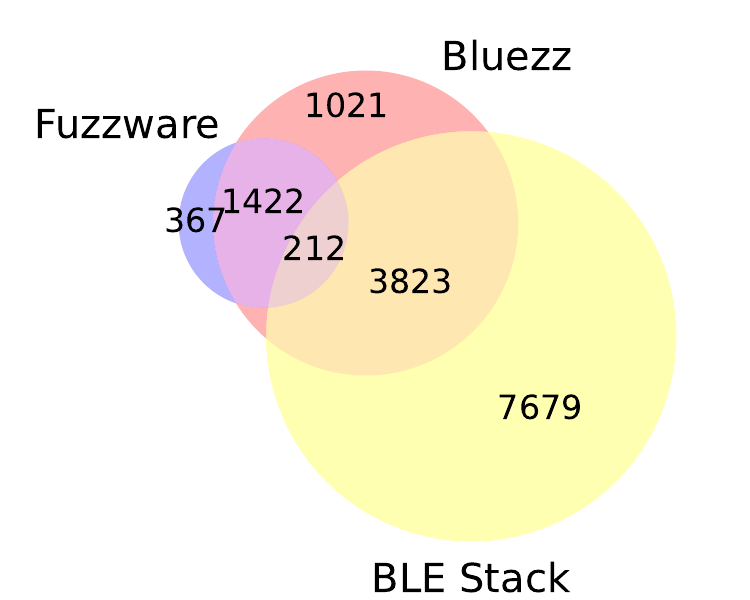}
        \caption{Nordic GATT Client}
    \end{subfigure}%
  \caption{Venn diagram for BLE stack-related basic blocks discovered by
    \system{} and Fuzzware.
    \system{} is denoted in red, Fuzzware in blue, and the overall basic blocks
    in the firmware related to the BLE stack in yellow.
    }
  \label{fig:venncov}
\end{figure*}


The coverage results are shown in~\autoref{fig:stackcov}. The benefit of
\system{} is evident: it outperforms Fuzzware and Hoedur across all 18 targets.
Using the arithmetic mean of per-target relative stack-coverage improvement,
\system{} achieves 4.9$\times$ higher stack-related coverage than Fuzzware and
5.9$\times$ higher coverage than Hoedur on average.
Further investigation reveals that Fuzzware and Hoedur suffer from limited 
emulation stability when handling fuzzed values, which constrains the firmware 
to processing only a small number of packet inputs.
As also observed in PEmu~\cite{bley2025protocol}, their network-related 
coverage is largely confined to the BLE advertising phase (i.e., advertising, 
and scanning), preventing the establishment of a full connection and thereby 
restricting access to higher-level application logic.
In contrast, \system{} provides stable and high-fidelity emulation, enabling 
the firmware to process packets sequentially.
For scanner-type targets, however, the coverage curves remain relatively flat 
across all tools, as scanners primarily receive and parse incoming packets 
without triggering extensive control flows.
Consequently, \system{}'s coverage growth plateaus once the packet-handling 
routines are fully exercised for the scanner-type targets.
In contrast, Fuzzware and Hoedur show minor fluctuations and slight increases 
in their coverage curves, as they occasionally trigger random interrupts.

The internal baseline without input models further clarifies the role of
structured input modeling. \system{} without input models quickly plateaus after
an initial coverage increase. So rehosting alone is insufficient for complex BLE firmware. Without the input model, the fuzzer struggles to further explore 
the protocol logic due to the lack of grammar and state dependencies.
This indicates that RPM's ECA rules are important for handling peripheral
modeling and for restricting fuzzed data to the packet-reception interface,
rather than exposing internal hardware interfaces as arbitrary fuzzing inputs.


To further understand if these fuzzers find the same set of basic blocks, we 
analyze their coverage overlap.
~\autoref{fig:venncov} illustrate the Venn diagrams comparing 
\system{} and Fuzzware. The remaining results are similar and appear in
Appendix~\ref{app:venncov}.
The results indicate that the three fuzzers discover largely disjoint sets of 
basic blocks. 
\system{} achieves higher coverage within BLE-stack–related regions, whereas 
Fuzzware tends to explore more non–BLE-stack code paths.
Further examination shows that the functions associated with the triggered 
basic blocks span all major BLE protocol phases, indicating that \system{} is 
able to exercise code paths across the full protocol lifecycle, even though not 
all functions within each phase are covered.




\subsection{Vulnerability Detection Analysis (RQ3)}

We conduct crash analysis and count the unique crashes. 
To deduplicate crashes, we hash the last three stack frames of each backtrace,
consistent with standard fuzzing practice~\cite{klees2018evaluating}.
Next, we replay the deduplicated traces to determine whether the crash occurs 
after the RADIO peripheral receives packets; such cases are counted as BLE 
stack–related unique crashes.
Finally, we attempt to validate crashes by replaying the corresponding packets
on development boards and performing root cause analysis using the emulation
environment.

Across all 18 samples, \system{} uncovered 10 unique crashes. After manual
root-cause analysis, we confirmed five previously unknown vulnerabilities, as
listed in Table~\ref{tab:bug-overview}. We have responsibly disclosed all of
them to the affected vendors. At the time of writing, one
vulnerability has been fixed.
Notably, all five vulnerabilities were triggered in the post-connection stage.
This highlights \system{}'s ability to drive BLE firmware beyond shallow
advertising and scanning logic and explore deeper, security-critical protocol
states that are difficult to reach
In contrast, baseline systems such as Fuzzware and Hoedur report many stack-unrelated crashes.
Fuzzware discovered 2, 3, and 1,215 crashes on the NimBLE, Zephyr, and Nordic samples, respectively, while Hoedur found 47, 1,817, and 38.
None of these crashes are attributable to BLE-stack code; they arise from malformed or emulation-specific inputs that cannot be replayed on real hardware. 
Some may originate from peripheral drivers or other non-BLE logic, but fully triaging every crash is infeasible under such volume.
This discrepancy also highlights the benefit of \system{}'s high-fidelity peripheral 
modeling, which avoids such crashes and focus on the OTA attack vector.

\begin{table}[t]
  \caption{Previously unknown vulnerabilities by \system{}.}
  \label{tab:bug-overview}
  \centering
  \resizebox{0.9\linewidth}{!}{
  \begin{tabular}{@{}p{0.22\linewidth}p{0.62\linewidth}p{0.10\linewidth}@{}}
    \toprule
    Target & Description & Status$^{*}$ \\ \midrule
    Zephyr Basic Peripheral &
    A disconnect-time race between thread context and mayfly / ISR context in the LLCP \texttt{proc\_ctx} release path can corrupt the free list in \texttt{mem\_release()} and cause a double free. &
    R \\
    Zephyr GATT Server &
    An all-zero \texttt{LL\_CHANNEL\_MAP\_IND} is accepted without validation, later causing an out-of-range channel selection and an assertion in \texttt{lll\_chan\_set()}. &
    R \\
    Zephyr GATT Server &
    A crafted post-connection PDU with manipulated \texttt{NESN} can trigger invalid ACK processing before TX state is initialized, leading to inconsistent bookkeeping and an assertion. &
    R \\
    Nimble GATT Server &
    A \texttt{LL\_CONNECTION\_UPDATE\_IND} with an invalid latency field is accepted and later causes 16-bit wraparound in connection scheduling, triggering \texttt{BLE\_LL\_ASSERT}. &
    R \\
    Nimble GATT Server &
    A zeroed \texttt{LL\_CHANNEL\_MAP\_IND} is copied into the pending channel map without validation; when activated, it reaches \texttt{ble\_ll\_utils\_remapped\_channel()} and triggers an assertion. &
    R/F \\ \bottomrule
  \end{tabular}%
  }

  \raggedright
  \footnotesize{$^{*}$R: Reported, F: Fixed.}
\end{table}

\subsection{Generality of RPM (RQ4)}
\begin{table}[]
    \caption{Summary of ECA-based peripheral modeling effort across MCU
    platforms. The effort is a one-time cost for modeling MCU peripherals and
    can be reused by firmware targets built on the same platform.}
  \label{tab:ecarules-effort}
  \centering
  \resizebox{\linewidth}{!}{%
  \begin{tabular}{@{}llll@{}}
  \toprule
  MCU platform & \#Peripherals & \#Rules for all peripherals & Estimated time \\ \midrule
  Nordic nRF52840 & 9 & 91 & $\sim$12.5h \\
  STM32F4         & 7 & 49 & $\sim$ 4.0h \\ \bottomrule
  \end{tabular}%
  }
\end{table}

\begin{figure}[t!]
	\centering
    \begin{subfigure}{0.24\textwidth}
        \centering
        \includegraphics[width=0.8\textwidth]{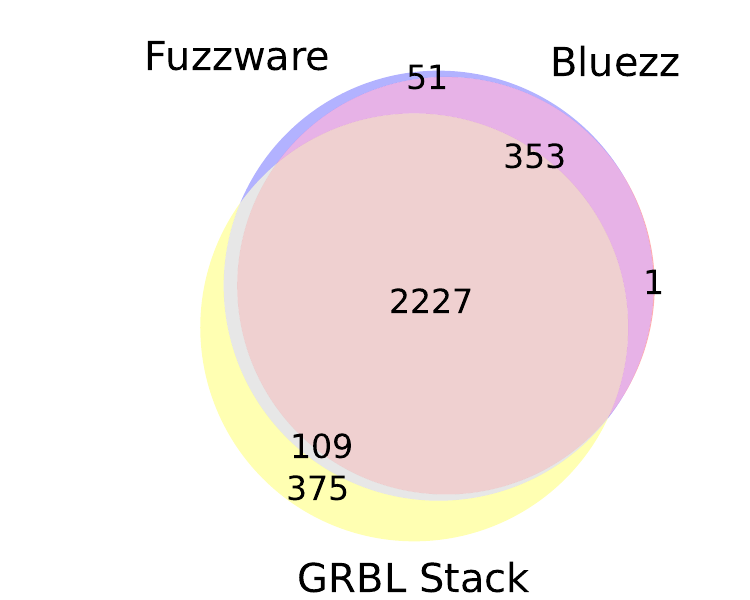}
        \caption{CNC Firmware}
    \end{subfigure}%
    \begin{subfigure}{0.24\textwidth}
        \centering
        \includegraphics[width=0.8\textwidth]{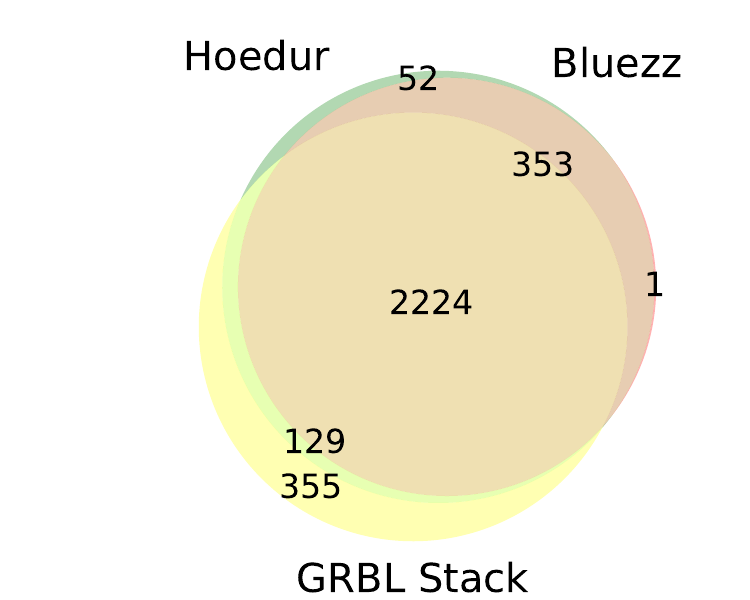}
        \caption{CNC Firmware}
    \end{subfigure}%
  \caption{Venn diagram for GRBL stack-related basic blocks discovered by
    \system{}, Fuzzware, and Hoedur.
    \system{} is denoted in red, Fuzzware in blue, Hoedur in green, and the overall basic blocks
    in the firmware related to the BLE stack in yellow.
    }
  \label{fig:cnnvenncov}
\end{figure}
We further extend \system{} to Grbl~\cite{grbl}, a real-world serial-protocol firmware for Computer Numerical Control (CNC) milling. 
Grbl is a representative non-BLE embedded target and is widely used in real world CNC controllers.
It has also been adopted as
a benchmark firmware in prior firmware rehosting and fuzzing
work~\cite{feng2020p2im, wang2025aidfuzzer, scharnowski2022fuzzware,
hofhammer2024surgeon, scharnowski2023hoedur}.
We run \system{}, Fuzzware, and Hoedur on Grbl for 24 hours across five
campaigns. The results are shown in~\autoref{fig:cnnvenncov}. All three fuzzers
achieve comparable code coverage, reaching over 70\% coverage and rediscovering
two known bugs. This suggests that RPM can be extended beyond BLE firmware while
maintaining fuzzing effectiveness on a widely used non-wireless benchmark.

We also estimate the manual effort required to support different MCU platforms,
as shown in Table~\ref{tab:ecarules-effort}. Modeling Nordic nRF52840 requires 12.5 hours for 9 peripherals
and 91 ECA rules, mainly due to BLE-specific peripherals such as RADIO, TIMER,
RTC, PPI, and CCM. The RADIO peripheral alone requires 38 rules to model packet
reception, transmission, timing, and interrupt behavior. In contrast, supporting
STM32F4 for Grbl requires 4.0 hours for 7 peripherals and 49 rules, as Grbl
mainly interacts with simpler peripherals such as UART, timers, and GPIOs.
Although the two MCU families expose different register designs, RPM remains
applicable. For example, Nordic peripherals often use separate registers for
tasks, events, and configurations, whereas STM32 peripherals may encode control
and status information in different bit fields of the same register. Despite
these differences, the same RPM modeling strategy and action primitives are
sufficient to capture the required peripheral behavior.
Importantly, this modeling effort is incurred per MCU platform rather than per
firmware target. Once the peripheral models are implemented, they can be reused
for other firmware running on the same MCU platform. These results show that RPM
can generalize beyond BLE firmware with manageable manual effort
Furthermore, the effort in this section need not remain fully manual. A key
benefit of the RPM paradigm is that it expresses peripheral behavior through
structured, declarative Event--Condition--Action rules, making the modeling
process amenable to agent assistance. We experimented with an agent-assisted
workflow in which an agent extracts peripheral behavior from hardware
documentation and proposes the corresponding ECA rules, while a human analyst
reviews and validates the resulting models. For the Nordic nRF52840 platform,
this workflow reduced the time required to construct and validate the 91 rules
from approximately 12.5 hours to approximately 2 hours. This result not only
shows that agents can substantially reduce the upfront modeling cost, but also
highlights the practical value and extensibility of RPM's structured modeling
paradigm, while retaining human oversight for hardware-specific semantics.
\section{Discussion}

\noindent\textbf{Lessons learnt.} Our evaluation suggests three main lessons for
rehosting-based analysis of embedded BLE firmware. First, faithfully modeling
\emph{reactive peripheral semantics} is not merely an engineering detail, but a
prerequisite for meaningful security testing. In BLE stacks, progression into
attacker-relevant states depends on precise causal coordination among MMIO
effects, interrupts, DMA activity, and implicit peripheral state transitions.
Even small deviations in these event chains can stall execution, prevent valid
connection progress, or drive the firmware into semantically invalid states.
Second, the practically relevant BLE attack surface lies primarily in
\emph{connected and post-connection states}, rather than in early advertising
or scanning logic alone. Existing fuzzers that remain confined to shallow
protocol phases leave much of the real attack surface unexplored. Our results
show that reaching connection management, control-PDU handling, and GATT
processing requires both faithful peripheral behavior and input models that
respect protocol state and packet structure.
Third, the vulnerabilities we found indicate that many BLE stacks remain
fragile in their handling of \emph{stateful control traffic}. Several of the
identified flaws stem from insufficient validation of control-PDU parameters or
from unsafe assumptions about state transitions after a malformed packet has
been accepted. This suggests that security weaknesses in embedded BLE stacks
often arise not from isolated parsing bugs alone, but from the interaction
between malformed inputs, delayed protocol effects, and peripheral-driven
execution over subsequent connection events.

\noindent\textbf{Limitations and future work.}
While \system{} demonstrates the effectiveness of RPM for complex peripheral
modeling and realistic attack-surface exploration, several limitations remain.
First, parts of the current workflow still require effort, including
understanding protocol and hardware specifications, constructing peripheral
models, and preparing use-case drivers. In practice, ECA rules can be
generated either by human analysts or by an agent-based workflow. We have
already evaluated the latter and found that an agent can generate effective
rules for our models, suggesting that rule construction need not be a fully
manual process. Nevertheless, reliably validating generated rules and
preparing drivers still require human oversight, especially for undocumented
or hardware-specific behavior. Further automating this pipeline could reduce
the cost of supporting new peripherals and improve the scalability of RPM.
Second, we plan to extend \system{} to more real world firmware samples, 
SoC vendors, and protocol-driven embedded
systems, in order to better assess the generality of RPM and further support
firmware analysis across a wider range of platforms.

\section{Related Work}
\textbf{Bluetooth security analysis.} Bluetooth security has been extensively 
studied over the years \cite{wu2024sok}. Prior research has uncovered design 
flaws in BLE-based tracking systems~\cite{wu2024finding,yu2024security}, 
authentication mechanisms~\cite{antonioli2020bias}, 
and pairing protocols~\cite{von2021method} 
through manual analysis and formal verification~\cite{wu2022formal}. 
In parallel, several studies have explored automated approaches for detecting 
memory safety vulnerabilities in Bluetooth implementations. These efforts 
include: reverse engineering of Bluetooth firmware~\cite{wen2020firmxray}, 
which suffers from limited testing coverage; top-down 
fuzzing~\cite{heinze2020toothpicker}, which overlooks OTA threat 
models; and on-device 
fuzzing~\cite{garbelini2022braktooth,karim2023blediff,garbelini2020sweyntooth}, 
which tends to be slow, lacks deep introspection, and does not scale well.
The closest to our work is Jan's work \cite{ruge2020frankenstein}, 
which rehosted the BLE firmware and testing them. which rehosts BLE firmware 
for fuzz testing. However, their approach relies on extracting firmware 
snapshots using vendor-specific debugging features available only on Broadcom 
and Cypress chips, limiting its scalability. Additionally, their testing 
is constrained to a small set of initial functions available before device
pairing. In comparison, we are the first to unlock full-stack
BLE firmware testing.

\noindent\textbf{Rehosting-based firmware testing.} Rehosting has proven 
valuable for testing low-end firmware including firmware 
incorporating trusted execution environments~\cite{harrison2020partemu}, 
cellular baseband firmware~\cite{hernandez2022firmwire}, and Linux-based router 
firmware~\cite{zheng2019firm,chen2016towards}. More recently, researchers have 
explored rehosting cortex-m firmware, which is the typical runtime 
environment for most BLE firmware. Due to the tight coupling between such 
firmware and underlying hardware, existing work focuses on modeling peripheral 
behavior through Memory-Mapped I/O (MMIO) 
\cite{feng2020p2im,mera2021dice,scharnowski2022fuzzware,scharnowski2023hoedur,
li2024basemirror} and high-level hardware abstraction layer (HAL) 
\cite{clements2020halucinator,hofhammer2024surgeon,seidel2023forming}. However, 
these approaches often suffer from limited fidelity or scalability. 
Specifically, HAL-based rehosting requires expert knowledge to manually 
identify relevant functions and implement accurate replacements, which is 
time-consuming and error-prone. Furthermore, such abstraction may overlook 
low-level hardware interactions critical for comprehensive analysis. MMIO-based 
rehosting, on the other hand, relies on symbolic execution 
\cite{feng2020p2im,scharnowski2022fuzzware} and board specifications
\cite{zhou2022your}
to emulate peripheral behavior, but these techniques may fail 
to capture complex or timing-sensitive interactions, particularly in BLE 
RADIO peripherals. By contrast, we utilize an event-condition-action 
modeling 
to define peripheral behaviors with improved fidelity.


\section{Conclusion}
Embedded BLE firmware remains difficult to analyze because it is typically
closed-source, tightly coupled with hardware, and driven by reactive
peripheral behaviors that are hard to reproduce faithfully in emulation. 
Existing rehosting approaches often fail to reach attacker-relevant protocol states and instead expose unrealistic attack surfaces through generic MMIO values or interrupt schedules.
We presented Reactive Peripheral Modeling (RPM), a new modeling paradigm that
uses an event-condition-action abstraction to represent peripherals as reactive
systems, and instantiated it in \system{}.
By capturing causal event chains, implicit peripheral state
transitions, and cross-peripheral coordination, \system{} enables semantically
valid firmware execution and aligns fuzzing with practical over-the-air attack
vectors.
Our extensive evaluation on BLE firmware samples shows that \system{} systematically exercises BLE stacks beyond
advertising and scanning into connection and post-connection states.
In doing so, \system{} uncovers five previously unknown vulnerabilities
located in post-connection states.
We further show that RPM is
not limited to BLE by extending it to protocol-driven firmware from another
MCU family with manageable manual effort.
Overall, \system{} demonstrates that faithful modeling of reactive peripheral
semantics is key to unlocking realistic and scalable embedded firmware fuzzing.
RPM and \system{} move rehosting-based security testing closer to
the conditions under which embedded devices are actually attacked, making it
more practical and security-relevant for modern embedded systems.



\bibliographystyle{ACM-Reference-Format}

\bibliography{ref}

\appendix 
\section{Open Science} 
We support open and reproducible research and release all the artifacts to
enable reproduction.
Our anonymous artifact repository is available at \url{https://anonymous.4open.science/r/Bluezz-A6F6/}.
Upon acceptance, we will publicly release the following artifacts:

\begin{itemize}
\item \textbf{Source code.} The implementation of \system{}, including the core analysis code and the generated peripheral models.

\item \textbf{Experiment scripts and configurations.}
Scripts for running the experiments, together with the configuration files used to produce the reported results.

\item \textbf{Benchmarks and datasets.} The 18 BLE samples used in our evaluation, together with their input models. We also provide the source code and build artifacts for these BLE samples.

\item \textbf{Documentation.} A README describing the environment requirements, installation steps, expected runtime, and commands for reproducing the key experiments.
\end{itemize}

\section{Ethical Considerations} 


\noindent\textbf{Stakeholders.}
This work involves four main stakeholder groups.
First, BLE firmware users may be affected if vulnerabilities in deployed
firmware are exploited before fixes are available.
Second, BLE firmware developers and SDK maintainers may need to
invest effort in triaging reported bugs, developing patches, and improving their
testing pipelines.
Third, security researchers can benefit from the methodology and
artifacts introduced by \system{} for analyzing BLE firmware in a reproducible
setting.
Finally, the community benefits from a more robust BLE ecosystem, as BLE is
widely used in consumer, industrial, and medical IoT devices.

\noindent\textbf{Impacts and Principles.}
We pay special attention to the potential ethical issues raised by this work.
Our study follows the principles of beneficence and respect for law and public
interest. All SDKs and firmware images used in our evaluation are obtained from
legitimate and publicly available sources, including official vendor repositories
and open-source projects such as Zephyr, NimBLE, and Nordic SDKs.
All fuzzing and emulation experiments are conducted in isolated environments.
Our experiments do not transmit over-the-air BLE packets and therefore do not
interfere with real-world BLE devices, networks, or users.
Moreover, we do not collect, inspect, or analyze personal data, device
identifiers, or user-generated traffic.

\noindent\textbf{Mitigations.}
The main potential risk is that vulnerability details could be misused against
unpatched BLE firmware. To reduce this risk, we follow responsible disclosure
practices. Newly discovered vulnerabilities are privately reported to the
corresponding maintainers or vendors before public disclosure. We avoid releasing weaponized exploits or instructions that would enable direct attacks against deployed devices. 
When necessary, we provide only the information required for
maintainers to reproduce and fix the issue.

\noindent\textbf{Decision.}
We decide to conduct and publish this research because its defensive benefits
outweigh the potential risks. The goal of \system{} is to improve the reliability
and security of BLE firmware through systematic testing, reproducible evaluation,
and responsible reporting. By helping developers identify and fix bugs before
they can be exploited in practice, this work contributes to the long-term
hardening of the BLE firmware ecosystem.


\section{BLE Protocol Stack}
\begin{figure}[htbp]
  \centering
  \includegraphics[width=0.8\linewidth]{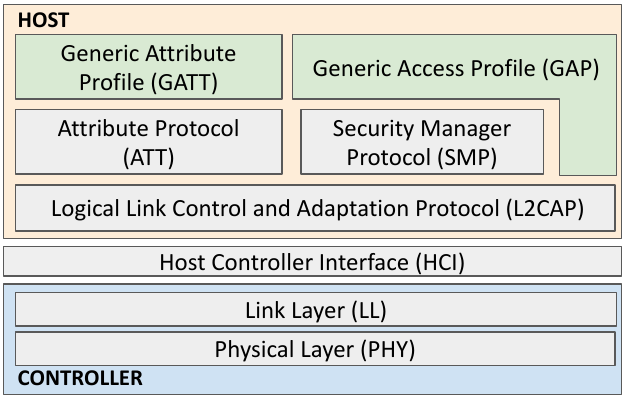} 
  \caption{The structure of BLE stacks. 
  }
  \label{fig:blestacks} 
\end{figure}

We present a summary of the BLE stacks and representative packet structures.
As shown in ~\autoref{fig:blestacks}, the BLE stack is 
typically divided into 
two main layers, including the controller and the host.
At the lower level, the BLE controller includes the Physical Layer (PHY) and
the Link Layer (LL), which are responsible for radio transmission and low-level 
link management, including advertising, scanning, and connection establishment.
Above that, the host includes Logical Link Control and Adaptation Protocol 
(L2CAP), Attribute Protocol (ATT), and Security Manager Protocol (SMP).
L2CAP manages data multiplexing and segmentation/reassembly, serving as a 
foundation for both ATT and SMP.
ATT enables clients to read and write attributes exposed by the server, 
forming the basis of BLE data communication, while SMP handles pairing and 
key distribution to secure the connection.
On top of these protocols, the Generic Attribute Profile (GATT) defines 
how devices exchange structured data using the ATT protocol, while the Generic 
Access Profile (GAP) governs device discovery, connection, and security 
procedures, built upon SMP and L2CAP.
While protocols define how data is exchanged, profiles define what behaviors 
and roles are expected for interoperability.

\begin{figure}[tbp]
  \centering
  \includegraphics[width=\linewidth]{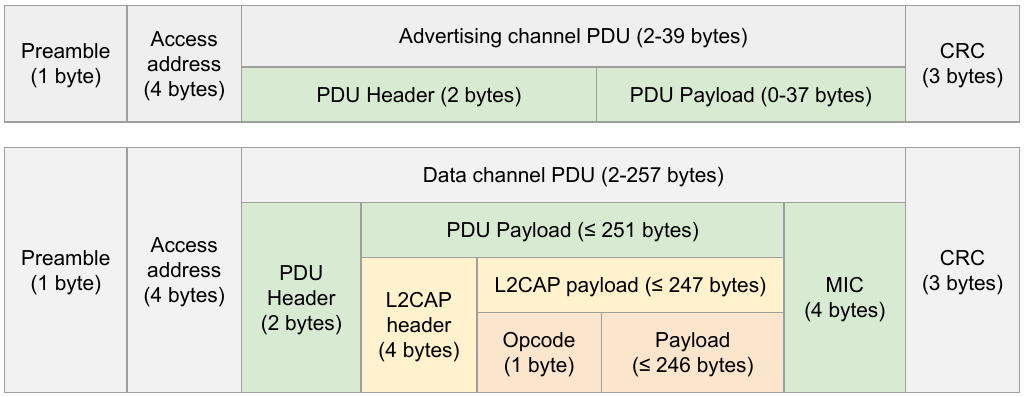} 
  \caption{Layered structure of BLE packets, including Link Layer advertising 
  packets and data packets.
  }
  \label{fig:blepackets} 
\end{figure}

Associated with the BLE protocol stack, packets are encapsulated 
across different layers.
As shown in~\autoref{fig:blepackets}, Link Layer packets are primarily 
categorized into advertising packets 
(e.g., \texttt{ADV\_IND}) and data packets used during active connections.
Link Layer data packets encapsulate L2CAP packets, which either carry 
upper-layer protocol messages such as ATT for attribute access and SMP for 
security procedures, or contain L2CAP signaling messages used for link control 
operations such as connection parameter updates. 

\label{app:blepackets}

\section{ECA Action Primitives}
\label{app:ecatemplates}
We present the built-in action primitives used to support the execution of Event–Condition–Action (ECA) rules in our system.
Our declarative ECA rule specification organizes each rule around events,
conditions, and actions: events define when a rule is triggered, conditions
define the guard predicates, and actions are realized through reusable runtime
primitives.
This appendix therefore focuses on the rule components needed to understand the
specification format, with particular emphasis on the action primitives that
model MMIO updates, interrupts, timers, counters, DMA-like packet transfers,
and inter-peripheral event routing.
While the current implementation is sufficient for modeling common
peripherals (e.g., RADIO, TIMER, DMA, and IRQ controllers), the action layer
is extensible and can incorporate additional primitives to support other
embedded domains.



We offer thirteen types of action primitives.
\begin{itemize}
  \item \texttt{update\_field}: modify a register, flag, or bitfield.
    \begin{itemize}
      \item \texttt{set(val)}: assign a value.
      \item \texttt{add(val)}: increment by value.
      \item \texttt{decrease(val)}: decrement by value.
      \item \texttt{or(target)}: bitwise OR with another field or value.
      \item \texttt{and(target)}: bitwise AND with another field or value.
      \item \texttt{not(target)}: bitwise NOT or clear (depending on context).
    \end{itemize}
    \item \texttt{start\_timer}: start a timer with a given interval.  
    \item \texttt{stop\_timer}: stop a running timer.  
  
    \item \texttt{start\_counter}: start a counter with a given interval.  
    \item \texttt{stop\_counter}: stop a running counter.  
  
    \item \texttt{start\_rcv\_pkt}: write a buffer (typically a fuzzed data 
    chunk) of the specified length 
    from a given source address to a designated target address. 
    \item \texttt{stop\_rcv\_pkt}: stop the ongoing packet reception.  
  
    \item \texttt{start\_send\_pkt}: read a buffer of the specified length 
    from a given source address.  

    \item \texttt{stop\_send\_pkt}: stop the ongoing packet transmission.  
  
    \item \texttt{fire\_irq}: assert (raise) the interrupt line with the given 
    IRQ number; the emulator in \system{} then triggers the corresponding 
    interrupt handler in the firmware.  
  
    \item \texttt{enable\_irq}: enable the specified device interrupt source, 
    allowing the corresponding interrupt to be delivered to the firmware.  

    \item \texttt{disable\_irq}: disable the specified device interrupt 
    source, preventing the interrupt from being delivered to the firmware.  
  
    \item \texttt{events\_route\_to\_tasks}: route hardware \texttt{EVENTS} 
    to \texttt{TASKS} for on-chip event chaining.
\end{itemize}

\emph{Note:} All primitive actions accept parameters that can reference a 
register or bitfield name, a constant value, or a user-defined global variable 
for state sharing across rules.

\section{Covered Use Cases of BLE Firmware Samples}
\label{app:usecase}

\autoref{tab:usecases} presents the summary of use cases covered by \system{}.

\begin{table}[htbp]
  \caption{Covered use cases of BLE firmware samples}
  \label{tab:usecases}
  \resizebox{\linewidth}{!}{%
  \begin{tabular}{@{}ll@{}}
  \toprule
    Use case                                                 & Functionalities \\ \midrule
  \multirow{2}{*}{Scanner} & Passive scan    \\
                           & Active scan     \\ \midrule
  \begin{tabular}[c]{@{}l@{}}Central (Basic \\ connection\\ management)\end{tabular} &
    \begin{tabular}[c]{@{}l@{}}1. Scan and initialize a connection\\ 2. Receive features exchange\\ 3. Receive version exchange\\ 4. Receive connection parameter change\\ 5. Receive disconnect request\\ 6. Disconnect\end{tabular} \\\midrule
  \begin{tabular}[c]{@{}l@{}}Peripheral (Basic\\ connection\\ management)\end{tabular} &
    \begin{tabular}[c]{@{}l@{}}1. Advertise and receive a connection\\ 2. Receive a connection parameter update\\ 4. Receive features exchange\\ 5. Receive version exchange\\ 6. Receive connection parameter change\\ 7. Receive disconnect request\end{tabular} \\\midrule
  \begin{tabular}[c]{@{}l@{}}Central \\ (GATT client)\end{tabular} &
    \begin{tabular}[c]{@{}l@{}}1. Scan and initialize a connection\\ 2. Request MTU exchange\\ 3. Discover primary services and characteristics\\ 4. Request characteristic values/descriptors read/write\\ 5. Receive characteristic value notification / indications\end{tabular} \\\midrule
  \begin{tabular}[c]{@{}l@{}}Peripheral\\ (GATT server)\end{tabular} &
    \begin{tabular}[c]{@{}l@{}}1. Advertise and accept a connection\\ 2. Receive MTU exchange request\\ 3. Receive primary services and characteristics discovery\\ 4. Receive characteristic values/descriptors read/write request\end{tabular} \\ \bottomrule
  \end{tabular}%
  }
  \end{table}

\section{Extended Venn Diagrams for Stack-related Basic Blocks}
\label{app:venncov}

\begin{figure*}[t!]
	\centering
    \begin{subfigure}{0.24\textwidth}
        \centering
        \includegraphics[width=0.8\textwidth]{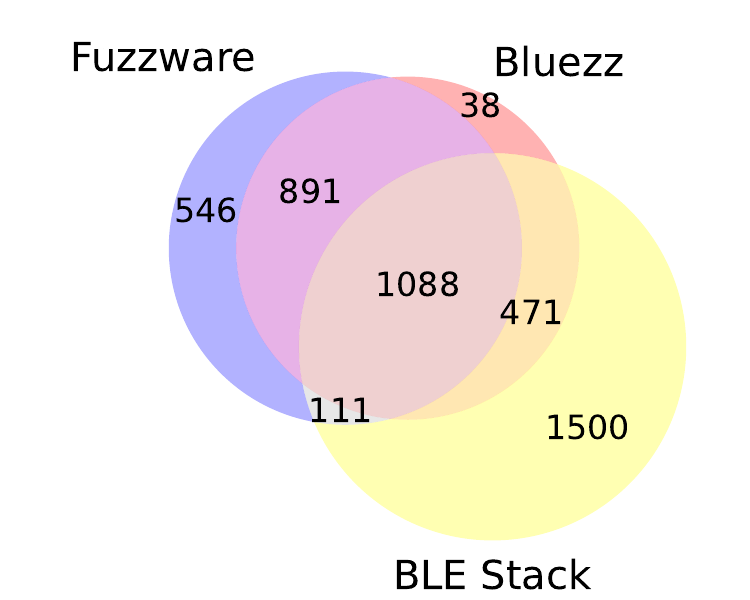}
        \caption{Nimble Active Scanner}
    \end{subfigure}%
    \begin{subfigure}{0.24\textwidth}
        \centering
        \includegraphics[width=0.8\textwidth]{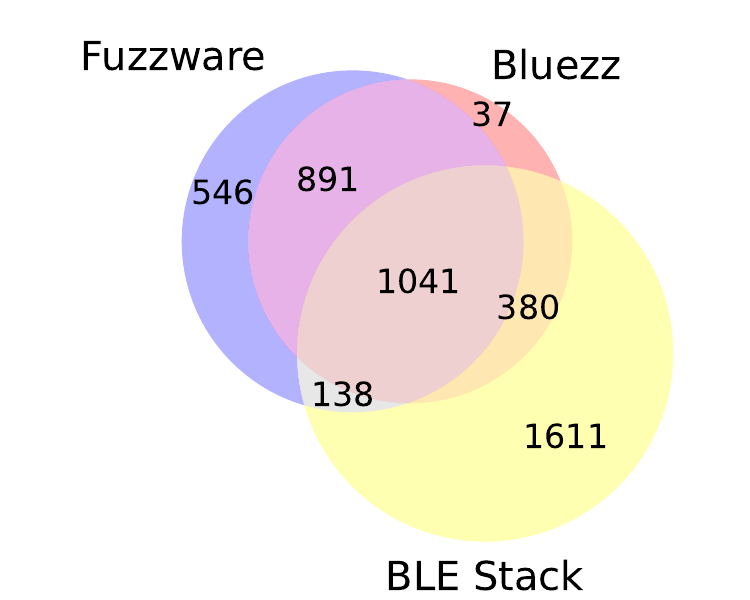}
        \caption{Nimble Passive Scanner}
    \end{subfigure}%
    \begin{subfigure}{0.24\textwidth}
        \centering
        \includegraphics[width=0.8\textwidth]{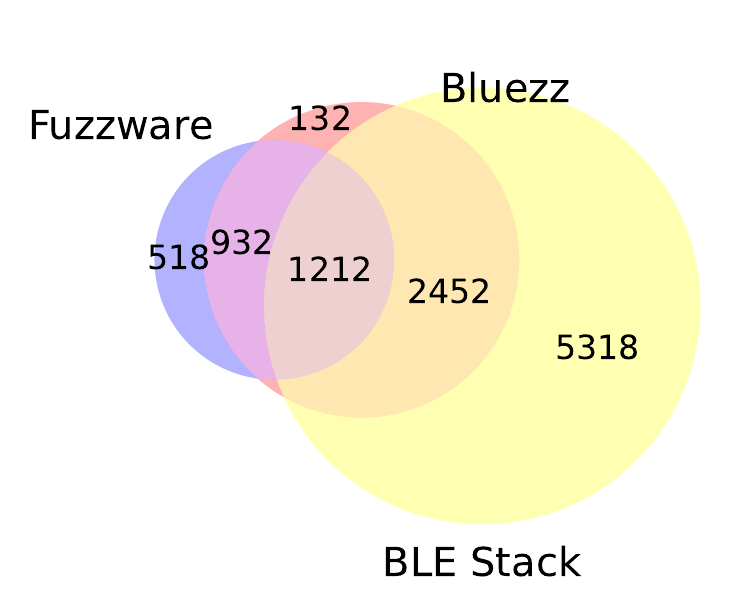}
        \caption{Nimble Basic Central}
    \end{subfigure}%
    \begin{subfigure}{0.24\textwidth}
        \centering
        \includegraphics[width=0.8\textwidth]{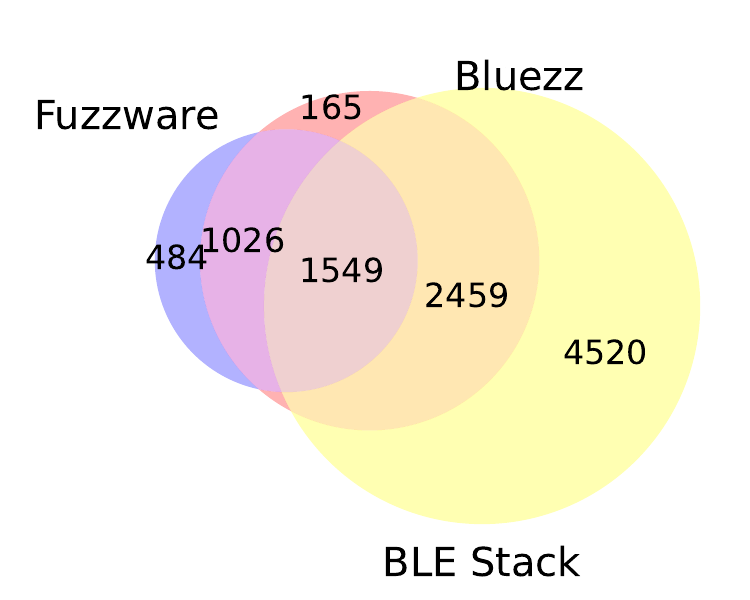}
        \caption{Nimble Basic Peripheral}
    \end{subfigure} \\
    \begin{subfigure}{0.24\textwidth}
        \centering
        \includegraphics[width=0.8\textwidth]{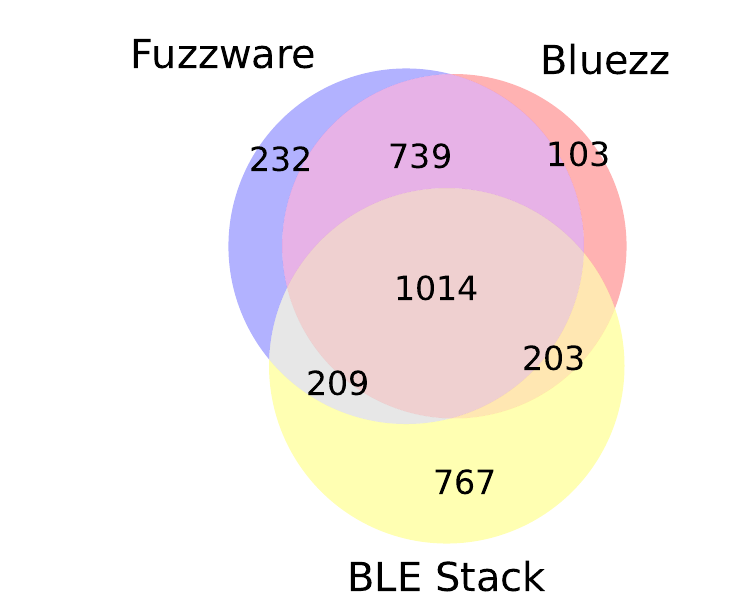}
        \caption{Zephyr Active Scanner}
    \end{subfigure}%
    \begin{subfigure}{0.24\textwidth}
        \centering
        \includegraphics[width=0.8\textwidth]{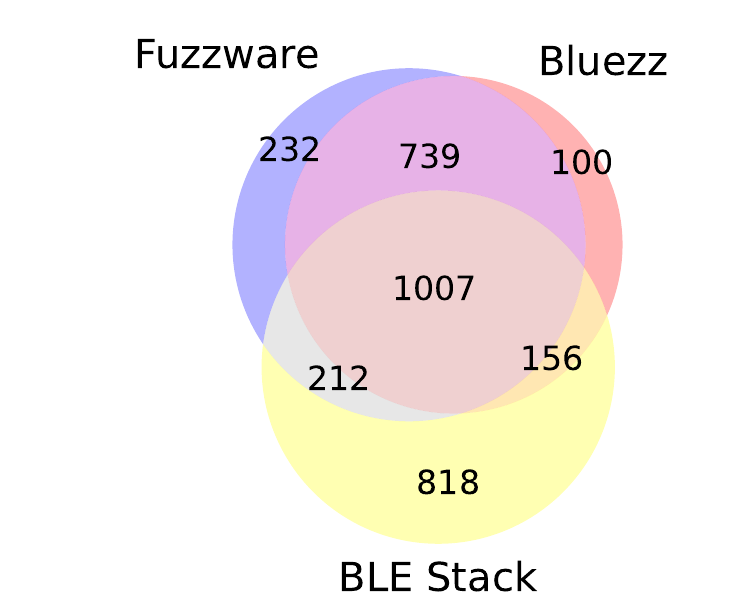}
        \caption{Zephyr Passive Scanner}
    \end{subfigure}%
    \begin{subfigure}{0.24\textwidth}
        \centering
        \includegraphics[width=0.8\textwidth]{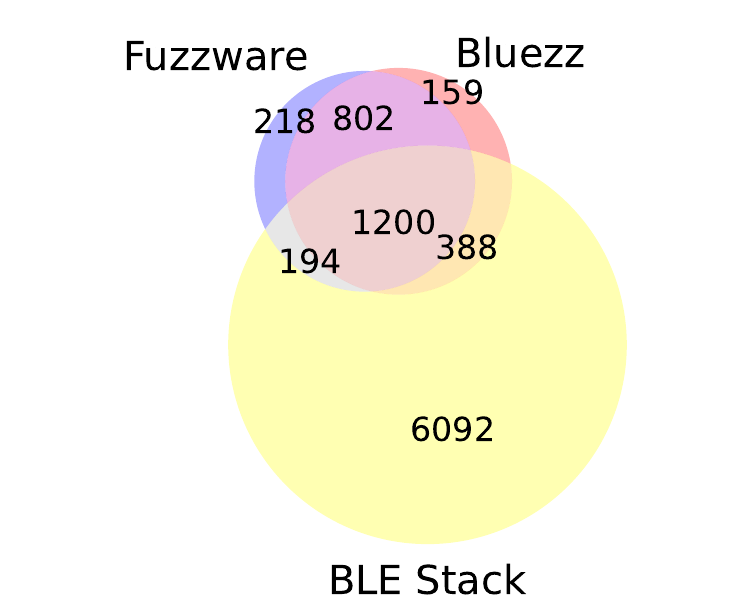}
        \caption{Zephyr Basic Central}
    \end{subfigure}%
    \begin{subfigure}{0.24\textwidth}
        \centering
        \includegraphics[width=0.8\textwidth]{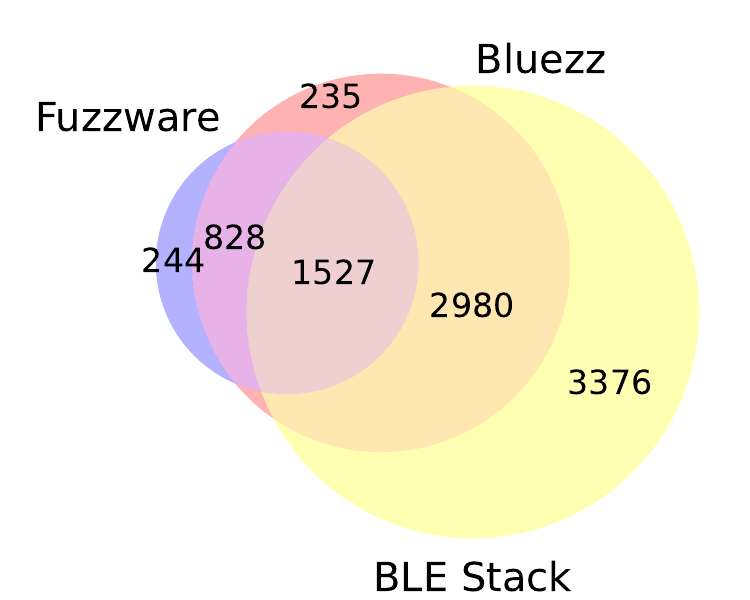}
        \caption{Zephyr Basic Peripheral}
    \end{subfigure}\\
    \begin{subfigure}{0.24\textwidth}
        \centering
        \includegraphics[width=0.8\textwidth]{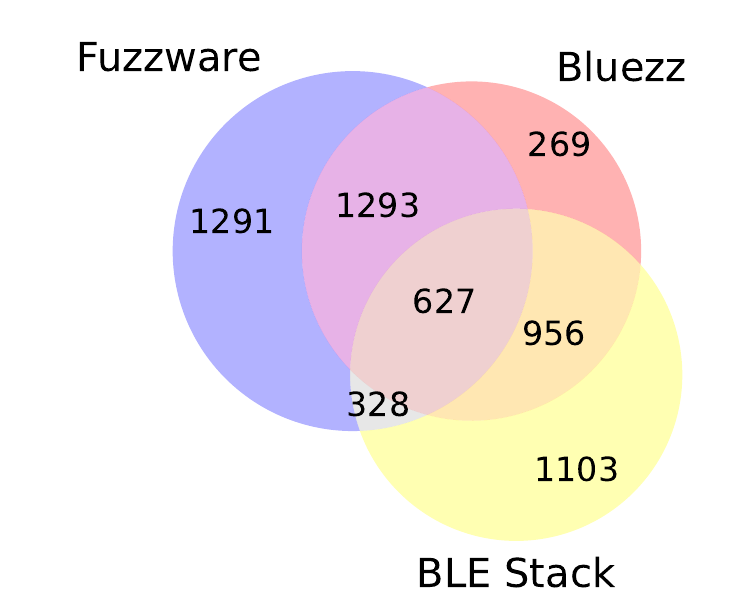}
        \caption{Nordic Active Scanner}
    \end{subfigure}%
    \begin{subfigure}{0.24\textwidth}
        \centering
        \includegraphics[width=0.8\textwidth]{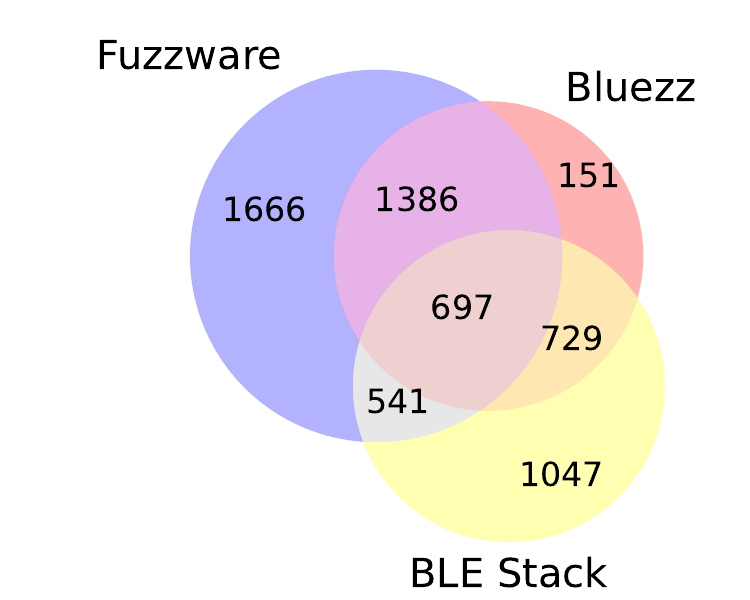}
        \caption{Nordic Passive Scanner}
    \end{subfigure}%
    \caption{Venn diagram for BLE stack-related basic blocks discovered by
        \system{} and Fuzzware. These are the diagrams for the targets not
        already presented in~\autoref{fig:venncov}.
        \system{} is denoted in red, Fuzzware in blue, and the overall basic blocks
        in the firmware related to the BLE stack in yellow.
        We omit Hoedur’s results here because they are highly similar to Fuzzware for these targets, and are excluded due to page-limit constraints.
        }
  \label{fig:app-venncov-fuzzware}
\end{figure*}

\autoref{fig:app-venncov-fuzzware} present
the extended Venn diagrams for all targets not presented in the evaluation in
\autoref{sec:evaluation}, \autoref{fig:venncov}.



\end{document}